%% file: main.tex
\documentclass[a4paper,11pt]{article}
\usepackage{jheppub}
\usepackage{amsmath}
\usepackage{amssymb}
\usepackage{graphicx}
\usepackage{booktabs}
\usepackage{multirow}
\usepackage{xcolor}
\usepackage{tabularx}
\usepackage{subcaption}
\usepackage{seqsplit}
\usepackage[compat=1.1.0]{tikz-feynman}
\usetikzlibrary{arrows.meta,calc,decorations.pathreplacing}

\title{Benchmarking Machine-Learning Top Taggers from JetClass to a Full-Simulation $Z'\to t\bar{t}$ Search: Transfer, Extrapolation, Sensitivity and Inference Cost}

\author[a]{Manisha Kumari,}
\author[a]{Amal Sarkar}

\affiliation[a]{School of Physical Sciences, Indian Institute of Technology Mandi,
Kamand, Himachal Pradesh 175075, India}

\emailAdd{manisha.kumari@cern.ch}
\emailAdd{amal.sarkar@cern.ch}

\abstract{Searches for heavy resonances decaying to \(t\bar{t}\) test physics beyond the Standard Model at the TeV scale, and their reach depends on how well boosted top jets are identified. Machine-learning top taggers are usually compared on fast-simulation samples without pileup and in a narrow jet-\(p_T\) window, so a benchmark ranking does not directly show how much each tagger adds to a real search. This study trains seven taggers under identical conditions on JetClass. They are a boosted decision tree, two image-based networks, ParticleNet, and three particle transformers (ParT, UParT, and GloParT). The taggers are then applied without retraining to a \(Z'\to t\bar{t}\) search with \(m_{Z'}\) between 1.2 and 2.5~TeV, using 2016 CMS Open Data simulated with Geant4 and realistic pileup. The overall benchmark order of the taggers is preserved. However, taggers with similar area under the ROC curve (AUC) differ by up to a factor of three in jet-level discovery significance. The particle transformers extend the mass reach of the search by \(400~\mathrm{GeV}\), and their lead grows with resonance mass. This gain is seen in fully merged top jets and is associated with stable performance at jet \(p_T\) above the training range. Pileup has only a minor effect on all taggers. A small convolutional network achieves competitive discrimination at a much lower computational cost. The gaps between tagger groups hold under retraining and within statistical uncertainties. The results show that a tagger must be chosen on its kinematic coverage, the working point of the analysis, its discovery sensitivity and its cost, and not on its benchmark score alone.}


\begin{document}
\maketitle
\flushbottom

\section{Introduction}
\label{sec:intro}

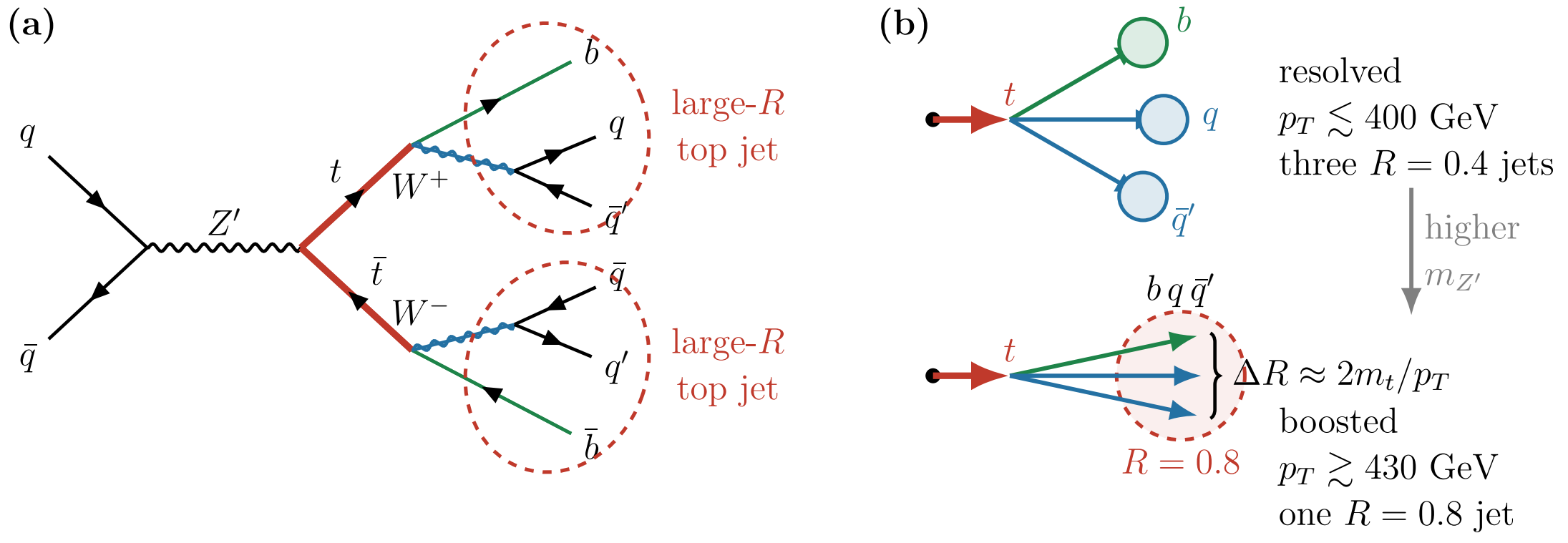
\begin{figure}[htbp]
\centering
\resizebox{\textwidth}{!}{\input{paper_figures/zprime_illustration_body}}
\caption{Schematic illustration of the signal. (a)~Production of a $Z'$ boson in $q\bar q$ annihilation and its decay to $t\bar t$, with both top quarks decaying hadronically and each decay contained in a large-radius jet. (b)~At low top $p_T$, the three decay quarks form separate jets, while above about $430~\mathrm{GeV}$ they are collimated within $\Delta R \approx 2m_t/p_T$ into a single jet of radius $R=0.8$.}
\label{fig:zprime-illustration}
\end{figure}

\begin{figure}[htbp]
\centering
\includegraphics[width=0.85\textwidth]{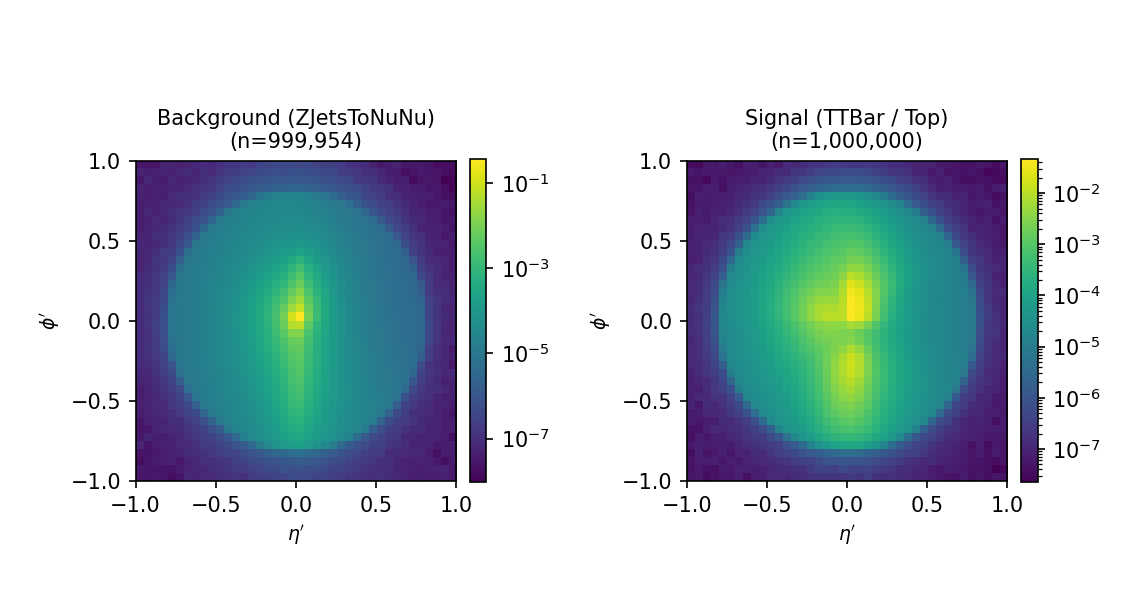}
\caption{Pixel-averaged jet image of the boosted-top quark, background (\texttt{QCD}) vs.\ signal
(\texttt{Hadronic Top}), in the rotated and aligned frame
described in Section~\ref{sec:reco}.}
\label{fig:jetclass-avg-image}
\end{figure}

\begin{figure}[htbp]
\centering
\includegraphics[width=0.95\textwidth]{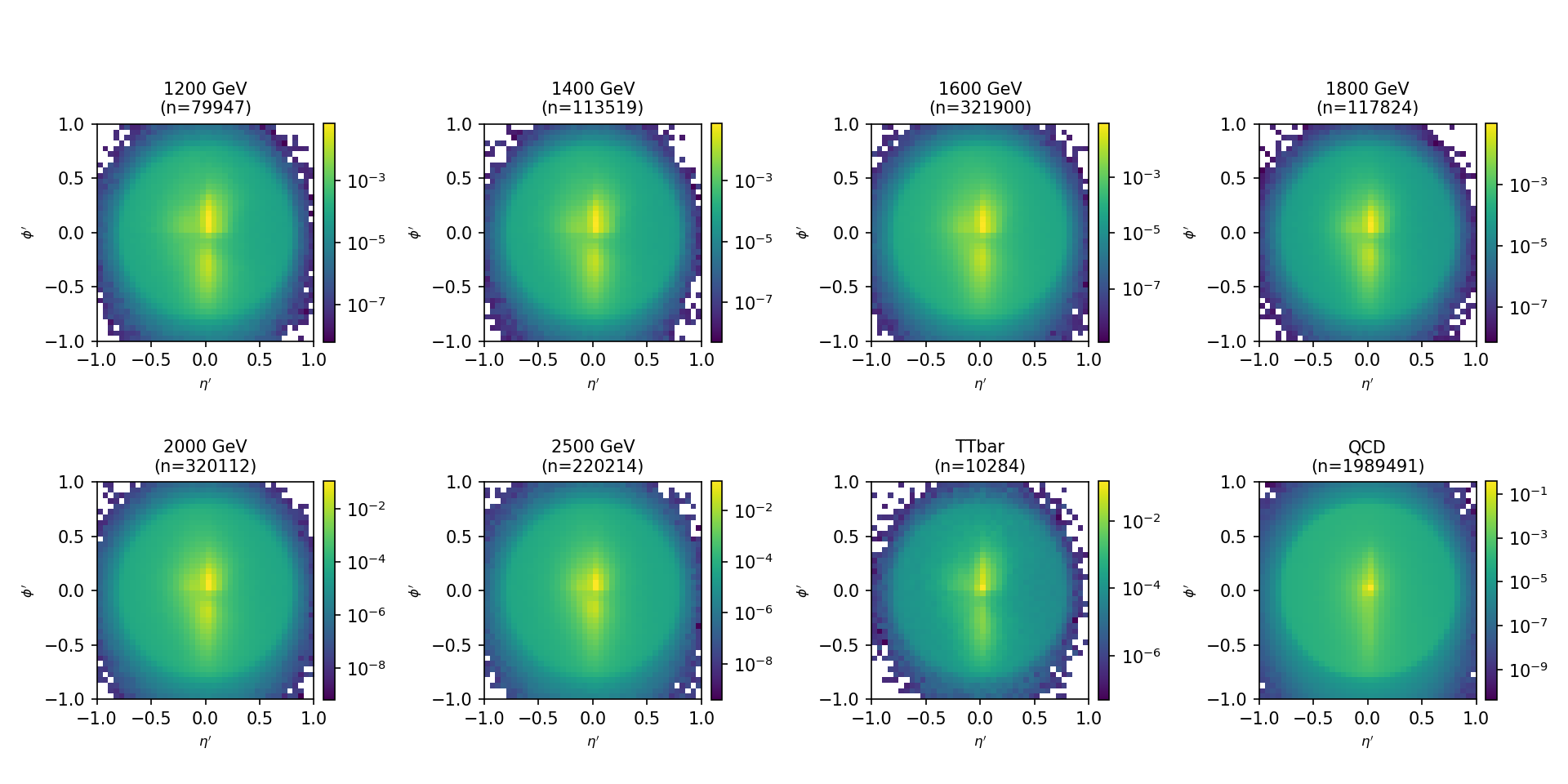}
\caption{Pixel-averaged signal jet images for each $Z'$ mass point
(1200--2500~GeV) alongside the SM $t\bar{t}$ and QCD backgrounds, in
the same frame as Figure~\ref{fig:jetclass-avg-image}.}
\label{fig:signal-vs-background-images}
\end{figure}

The top quark is the heaviest known elementary particle. Its mass of about $173~\mathrm{GeV}$ gives it a Yukawa coupling to the Higgs boson close to unity, which makes it a sensitive probe of electroweak symmetry breaking and a stringent test of Standard Model (SM) predictions. At the Large Hadron Collider (LHC) top quarks are produced in large numbers through SM processes, and they may also come from the decays of heavy particles predicted beyond the SM. Separating top quarks from the much larger background of light-quark and gluon jets is therefore central to both precision measurements and new-physics searches. Many extensions of the SM predict heavy resonances that couple strongly to third-generation quarks. A standard benchmark is the leptophobic topcolor $Z'$ boson, which decays mainly to $t\bar{t}$ \cite{ZprimeTopcolor} and is used in CMS searches for top-quark pair resonances \cite{CMSZprimeBenchmark}. Such a resonance would show up as a localized excess in the $t\bar{t}$ invariant-mass spectrum above a QCD multijet background that is many orders of magnitude larger. For resonance masses above about 1~TeV each top quark carries a transverse momentum of roughly $m_{Z'}/2$. The decay products of a hadronically decaying top quark are then strongly collimated. These are a $b$ quark and the two quarks from the $W$-boson decay. Their angular separation scales as $\Delta R \sim 2m_t/p_T$ and drops below the jet radius $R=0.8$ above about $430~\mathrm{GeV}$. In this boosted regime, all three decay products fall within a single large-radius jet. The jet then shows a three-prong substructure in which two of the prongs form a system close to the $W$-boson mass. Figure~\ref{fig:zprime-illustration} shows the production and decay of the resonance and the change from resolved to boosted topologies. Figure~\ref{fig:jetclass-avg-image} compares the average jet image of hadronic top quarks in the JetClass $t\bar{t}$ sample (see section~\ref{sec:samples}) with that of QCD jets. The top jets show three prongs, while the energy of QCD jets is concentrated in a single core. Figure~\ref{fig:signal-vs-background-images} makes the same comparison for simulated $Z'\to t\bar{t}$ signal and QCD background from CMS Open Data (see section~\ref{sec:samples}). The top-jet substructure becomes more collimated as the $Z'$ mass increases. Separating boosted top jets from QCD jets is known as \emph{top tagging}. It sets the background rejection of a heavy $t\bar{t}$ resonance search and so directly controls its sensitivity.

Top-tagging methods have developed through several generations, as reviewed in Ref.~\cite{JetSubstructureReview}. Early taggers fed hand-crafted substructure observables such as $N$-subjettiness \cite{Nsubjettiness}, energy correlation functions \cite{ECF}, and groomed jet mass \cite{SoftDrop} into boosted decision trees. These methods are easy to interpret, but they are limited to the observables chosen in advance. Convolutional neural networks applied to jet images \cite{JetImages, DeepTop} learned discriminating features directly from the energy deposits, and they were compared in community benchmarks \cite{TopTaggerLandscape}. Later methods work directly on the jet constituents and treat them as a permutation-invariant particle cloud \cite{DeepSets, PointNet}. ParticleNet \cite{ParticleNet} is a graph neural network that showed the strength of constituent-level learning. The Particle Transformer (ParT) \cite{ParT} was trained on JetClass and further improved tagging performance. In CMS, ParT followed DeepAK8 \cite{DeepAK8} as the leading approach, and the transformer design was later extended to the Unified Particle Transformer (UParT) \cite{UParT} and the Global Particle Transformer (GloParT) \cite{GloParT} for Run~3. These taggers are nonetheless usually compared on benchmark samples made with fast detector simulation, in a limited jet-$p_T$ range and without pileup \cite{ParticleNet, ParT, TopTaggerLandscape}. Their performance is summarised by the area under the receiver operating characteristic curve (AUC). A real resonance search works under different conditions. Its jet-$p_T$ spectrum shifts with the resonance mass, detector effects and pileup alter the reconstructed jets, and the useful working points often lie in the high-purity region. Two taggers with similar benchmark AUCs can therefore yield different search sensitivities. Computational cost also matters when inference runs over large event samples or within tight trigger latency limits.

Several earlier studies address parts of this problem. A study of an improved substructure tagger showed that a modest gain in top-tagging performance turns into a larger gain in $Z'$ search sensitivity \cite{Kasieczka:2015jma}. ATLAS compared constituent-based taggers, including ParticleNet, on fully simulated jets with pileup. That study also examined training-seed variation and inference time and found that the best-performing taggers exhibit larger systematic uncertainties \cite{ATLASAccuracyPrecision}. It did not include transformers and did not test the taggers in a resonance search. Other recent work transfers JetClass-pretrained models to full simulation through fine-tuning \cite{Mikuni:2025omnilearn, Bhimji:2025omnilearned, Schott:2026transfer}, and the robustness of ParT at high pileup has been studied for quark-gluon tagging \cite{Castillo:2025qg}. It has not yet been established whether benchmark rankings survive when taggers are applied without retraining to a detector-level resonance search. It is also open how their relative performance turns into discovery sensitivity across this range of architectures. This paper answers both questions with a matched comparison of seven existing taggers (Section~\ref{sec:models}) in a simulated $Z'\to t\bar{t}$ search using fully simulated CMS Open Data. The study measures how benchmark performance transfers to search conditions, how large the resulting differences in sensitivity are, and what each tagger costs to run. The aim is neither a new architecture nor a single best tagger. The aim is to show how tagger performance should be judged within the search that the tagger is meant to serve.

The seven taggers cover boosted decision trees (BDTs), image-based networks, ParticleNet, and three particle transformers, all of which are trained under common conditions on JetClass. They are evaluated in the $Z'\to t\bar{t}$ search at six resonance masses between $1.2$ and $2.5~\mathrm{TeV}$ using Geant4-simulated CMS Open Data with realistic pileup. The comparison covers discrimination and its dependence on resonance mass and jet $p_T$, the effect of pileup, discovery significance, and inference cost. The stability of the results is also checked under retraining and finite-sample statistical fluctuations. 
This study goes beyond previous comparisons in five ways. It tests the transfer of tagger performance from a public fast-simulation benchmark to full CMS detector simulation without retraining. It compares ParT, UParT and GloParT with ParticleNet, image-based networks and a BDT under the same training conditions. It examines performance both inside and above the jet-$p_T$ range of the training sample. It separates the effect of pileup from the shift in the jet-$p_T$ spectrum that comes with each resonance mass. Finally, it quantifies the consequences for discovery significance and mass reach together with the computational cost of each tagger.

Discovery significance is computed at the single-jet level and not from a fit to the $t\bar{t}$ invariant-mass spectrum. This choice isolates the contribution of the tagger to signal-background separation. It keeps the result free of effects from tagging the second top jet, from the invariant-mass resolution and from the choice of fit model. The significance is therefore a consistent basis for comparing taggers and not a full estimate of the sensitivity of an experimental search. The overall tagger ranking is largely preserved from the benchmark to full simulation, but the size of the performance differences changes substantially. Taggers with similar AUC differ by up to a factor of three in jet-level discovery significance at search-relevant working points. These conclusions hold for the implementations and training configurations used here. They do not rank the intrinsic capability of the underlying architectures.

The paper is organized as follows. Section~\ref{sec:samples} describes the training and application samples and how they differ. Section~\ref{sec:reco} covers jet reconstruction and the tagger inputs, and Section~\ref{sec:models} describes the architectures and training. Section~\ref{sec:results} presents the results. Section~\ref{sec:conclusion} summarises the findings and what they mean for tagger choice in boosted-top searches.

\section{Training and Application Datasets and Event Selections}
\label{sec:samples}

All taggers are trained on JetClass \cite{Qu2022JetClass}, a public dataset of simulated jets in ten categories,  generated with \textsc{MadGraph5\_aMC@NLO} \cite{MadGraph5} and \textsc{Pythia~8.2} \cite{Pythia8}, followed by CMS-like \textsc{Delphes} fast simulation \cite{Delphes, ParT}. The task is formulated as a binary classification problem, with \texttt{TTBar} jets as the signal and \texttt{ZJetsToNuNu} jets as the background. The jets are clustered using the anti-$k_T$ algorithm \cite{Cacciari2008, FastJet} with a radius parameter of $R=0.8$, using \textsc{Delphes} energy-flow objects and the JetClass selection requirements summarised in Table~\ref{tab:train_vs_apply}. The trained taggers are evaluated in a simulated $Z'\to t\bar{t}$ search using CMS Open Data simulation from the \texttt{\seqsplit{RunIISummer20UL16MiniAODv2}} campaign \cite{CMSOpenData}.  The signal consists of the leptophobic topcolour $Z'$ benchmark \cite{ZprimeTopcolor,CMSZprimeBenchmark} at six mass hypotheses: $m_{Z'}=1200$, $1400$, $1600$, $1800$, $2000$, and $2500~\mathrm{GeV}$. The samples correspond to the narrow-width scenario, with $\Gamma/m_{Z'}\approx1\%$, and are generated using \textsc{MadGraph5\_aMC@NLO} with MLM matching, interfaced to \textsc{Pythia~8} with the (\texttt{TuneCP2}) tune.
The dominant background is QCD multijet production, represented by five $H_T$-binned samples from the \texttt{\seqsplit{TuneCP5\_PSWeights\_13TeV-madgraph-pythia8}} campaign \cite{CP5Tune, NNPDF31}. All hadronic $t\bar{t}$ production, simulated using \texttt{\seqsplit{TTToHadronic\_TuneCP5\_13TeV-powheg-pythia8}}, provides an additional irreducible SM background. The leading large-radius jet is selected in each event.

\begin{table}[htbp]
\centering
\caption{Jet-$p_T$ windows per dataset in the $Z'\to t\bar{t}$ application sample part of the selection of Table~\ref{tab:train_vs_apply}. By design, every signal window extends below or above the $500$--$1000$~GeV training range of JetClass (Section~\ref{sec:results:domainshift}).}
\begin{tabular}{llcc}
\toprule
\textbf{Category} & \textbf{Dataset} & \textbf{Mass point (GeV)} & \textbf{$p_T$ window (GeV)} \\
\midrule
\multirow{6}{*}{Signal ($Z'\to t\bar{t}$)}
 & ZPrime\_M1200 & 1200 & 400--850 \\
 & ZPrime\_M1400 & 1400 & 450--950 \\
 & ZPrime\_M1600 & 1600 & 500--1050 \\
 & ZPrime\_M1800 & 1800 & 550--1150 \\
 & ZPrime\_M2000 & 2000 & 600--1300 \\
 & ZPrime\_M2500 & 2500 & 750--1600 \\
\midrule
\multirow{6}{*}{Background}
 & QCD\_HT500to700   & -- & 400--1600 \\
 & QCD\_HT700to1000  & -- & 400--1600 \\
 & QCD\_HT1000to1500 & -- & 400--1600 \\
 & QCD\_HT1500to2000 & -- & 400--1600 \\
 & QCD\_HT2000toInf  & -- & 400--1600 \\
 & TTToHadronic    & -- & 400--1600 \\
\bottomrule
\end{tabular}

\label{tab:pt_windows}
\end{table}

The event selection and jet-$p_T$ requirements are summarised in Table~\ref{tab:train_vs_apply}, which serves as the reference throughout this analysis, and Table~\ref{tab:pt_windows} lists the jet-$p_T$ windows used for each signal mass point and for the background samples. Since the characteristic transverse momentum of a top quark from a heavy resonance scales approximately as $p_T\approx m_{Z'}/2$, the signal window varies with the resonance mass. The background is selected over the full $400$--$1600~\mathrm{GeV}$ range, covering the union of the signal windows, reflecting the selection used in a mass-specific search. The leading large-radius jet satisfying the selection is retained in each event. Each selected jet is assigned the event weight
\begin{equation}
w = \frac{\sigma\,\mathcal{L}\,w_{\rm gen}}
{\sum w_{\rm gen}},
\label{eq}
\end{equation}
where $\sigma$ is the sample cross section, $\mathcal{L}$ is the integrated luminosity, $w_{\rm gen}$ is the generator weight, and the denominator is the sum of generator weights in the corresponding simulated sample. This normalization gives the expected event yield for the assumed integrated luminosity. Generator-level information is not used for event selection or as input to any tagger in the application samples. The collection \texttt{\seqsplit{prunedGenParticles}} is used only to study the signal composition in Section~\ref{sec:results}. ROC curves and AUC values are evaluated independently at each mass point, using the corresponding signal and background selections.

\begin{table}[htbp]
\centering
\caption{Jet selection and sample properties of the JetClass training
sample and the $Z'$ application sample.}
\small
\setlength{\tabcolsep}{4pt}
\begin{tabularx}{\textwidth}{@{}l>{\raggedright\arraybackslash}X>{\raggedright\arraybackslash}X@{}}
\toprule
\textbf{Property} & \textbf{JetClass (training)} & \textbf{MiniAOD (application)} \\
\midrule
Jet algorithm & anti-$k_T$, $R=0.8$, E-flow objects & anti-$k_T$, $R=0.8$, all PF candidates \\
Pileup mitigation & -- (no pileup) & none (no CHS or PUPPI) \\
Jets used per event & all jets passing selection & leading jet passing selection \\
$p_T$ (signal) & 500--1000~GeV & per mass point, Table~\ref{tab:pt_windows} \\
$p_T$ (background) & 500--1000~GeV & 400--1600~GeV (union of windows) \\
$|\eta|$ & $<2$ & $<2$ \\
Constituents & 50 leading (particle cloud), all (BDT, images) & 50 leading in $p_T$ (all taggers) \\
Simulation & \textsc{Delphes} fast simulation & Geant4 full simulation \\
Pileup & none & 2016 conditions ($\langle n_{\rm PV}\rangle=17.9$, max.\ 75--85) \\
Truth requirement & signal: all decay quarks within $\Delta R<0.8$ of jet axis & none \\
Label & generator-level class & none (tagger score only) \\
\bottomrule
\end{tabularx}

\label{tab:train_vs_apply}
\end{table}

\section{Jet Reconstruction and Tagger Inputs}
\label{sec:reco}

The JetClass training sample and the CMS Open Data application samples differ in detector simulation, pileup, kinematic coverage and constituent selection, as summarised in Table~\ref{tab:train_vs_apply}. JetClass uses \textsc{Delphes} fast simulation without pileup and keeps jets with $500$--$1000~\mathrm{GeV}$ in transverse momentum. The application samples use the full Geant4 simulation \cite{Geant4} of CMS with realistic pileup, and their jet-$p_T$ range extends from $400$ to $1600~\mathrm{GeV}$. These differences make it possible to test how tagger performance transfers beyond the conditions of the training sample.

The particle-cloud taggers need constituent-level four-momenta and particle identities. Large-radius jets are therefore reconstructed directly from particle-flow (PF) candidates in MiniAOD \cite{CMSParticleFlow,CMSDetector,CMSMiniAOD} with the anti-$k_T$ algorithm and $R=0.8$. No pileup mitigation algorithm, such as charged-hadron subtraction (CHS) or PUPPI, is applied. The JetClass jets contain no pileup, so this choice preserves the full difference between training and application conditions and provides a conservative test of the taggers' robustness. The leading jet that passes the selection in Table~\ref{tab:train_vs_apply} is kept, and its 50 highest-$p_T$ constituents are passed to each tagger. Nearly all selected jets have more than 50 constituents, so this limit removes the softer ones. It also creates a difference between training and application for the BDT and image-based taggers, which use all constituents that pass the JetClass selection in training. Track impact parameters are converted to the units used in JetClass. The number of reconstructed primary vertices $n_{\mathrm{PV}}$ is recorded as a measure of pileup \cite{CMSTracking} but is not given to any tagger as an input.

\textbf{Particle-cloud taggers:} The three transformers receive 17 features per constituent. These are seven kinematic variables ($\Delta\eta$, $\Delta\phi$, $\log p_T$, $\log E$, $\log(p_T/p_{T,\mathrm{jet}})$, $\log(E/E_{\mathrm{jet}})$, and $\Delta R$), six particle-identification features that encode charge and particle type, and four track-displacement variables ($d_0$, $d_z$, $\sigma_{d_0}$, and $\sigma_{d_z}$). The displacement variables are set to zero for neutral particles. Each constituent pair also carries four interaction features. These are the angular separation $\Delta$, the relative transverse momentum $k_T=p_{T,\min}\Delta$, the momentum-sharing fraction $z=p_{T,\min}/(p_{T,a}+p_{T,b})$ and the pairwise invariant mass squared $m^2=\max\!\left((E_a+E_b)^2-|\vec{p}_a+\vec{p}_b|^2,0\right)$. These pairwise features are not logarithmically transformed. ParticleNet receives seven kinematic node features and uses the $(\Delta\eta,\Delta\phi)$ coordinates to construct its $k$-nearest-neighbour graph. Unlike the transformers, it does not receive particle-identification or track-displacement features.

\textbf{Image-based taggers:} The CNN and ResNeXt process jet images constructed from the constituent transverse-momentum distribution in the $(\Delta\eta,\Delta\phi)$ plane. Each jet is centered on its $p_T$-weighted centroid, rotated so that its principal axis points vertically, and reflected so that the side carrying more $p_T$ lies in the top-right quadrant. Constituents are mapped to a grid spanning $\pm1.0$ in both coordinates, with each pixel weighted by the constituent $p_T$. The image is normalized to the total $p_T$ within the grid, and constituents outside the grid are consistently discarded during training and application. The CNN uses a single-channel $40\times40$ image. ResNeXt uses a $64\times64$ image replicated across three channels to accommodate an ImageNet-style backbone \cite{ImageNet}. The additional ResNeXt training seeds studied in Section~\ref{sec:results:uncertainty} use $40\times40$ images, and each checkpoint is evaluated with the image size used during its training.

\textbf{Boosted decision tree:} The BDT uses 19 high-level observables derived from subjets obtained by reclustering jet constituents with the anti-$k_T$ algorithm using $R=0.2$ and retaining subjets with $p_T>5~\mathrm{GeV}$. The observables are seven subjet-kinematic features, three energy-sharing ratios, and nine angular-spread variables. The kinematic features describe the subjet multiplicity and the maximum, mean, and standard deviation of subjet $p_T$ and mass. The energy-sharing variables quantify the total transverse momentum, the leading-subjet $p_T$ fraction, and the leading-subjet size fraction. The angular variables are the maximum, mean, and standard deviation of $\Delta R$, $|\Delta\eta|$, and $|\Delta\phi|$ between each subjet and the parent jet axis. The BDT is therefore a baseline. It uses neither the jet mass nor the $N$-subjettiness ratio $\tau_3/\tau_2$ \cite{Nsubjettiness}, and a fuller set of observables could improve its performance.

\section{Tagger Architectures and Training}
\label{sec:models}
\label{sec:training}
The seven taggers in Table~\ref{tab:models} follow three approaches to jet classification. These are high-level substructure observables, image-based representations and constituent-level learning. All models are trained from scratch on the same JetClass sample (see section~\ref{sec:samples}) for the same binary task, with $\texttt{TTBar}$ as signal and $\texttt{ZJetsToNuNu}$ as background. This matched setup allows their performance to be compared on equal terms. The BDT is built with XGBoost \cite{XGBoost} and uses high-level substructure observables. The image-based models are a compact convolutional neural network (CNN) and ResNeXt-50 \cite{ResNeXt,ResNetHe}. The CNN has three convolutional blocks followed by a classifier, while ResNeXt-50 is a much deeper computer-vision network applied to the same kind of jet images. Comparing the two tests whether the extra depth brings a real gain over the compact CNN. At the constituent level, ParticleNet \cite{ParticleNet} represents a jet as a graph of particles. Each particle learns its features from its neighbouring constituents through EdgeConv graph convolutions \cite{DynamicGraphCNN}. The three transformer-based models \cite{Transformer} ParT \cite{ParT}, UParT \cite{UParT} and GloParT \cite{GloParT} use attention to learn the relations between jet constituents. ParT adds pairwise particle features to its attention and builds the jet representation through a learned class token. UParT replaces the class-token readout with an average over the constituent representations. GloParT retains the class token but shifts layer normalization \cite{LayerNorm} to a different position. The three transformers thus share a common backbone and differ mainly in their readouts or normalizations.

The neural networks use a 70\%/15\%/15\% split of the JetClass sample into training, validation and test sets, and the BDT uses a 60\%/20\%/20\% split. For each model, the checkpoint with the lowest validation loss is kept. Table~\ref{tab:training} lists the optimizer, learning rate, batch size, and number of epochs. All neural networks are written in PyTorch \cite{PyTorch}. The analysis and plots use scikit-learn \cite{ScikitLearn}, NumPy \cite{NumPy} and Matplotlib \cite{Matplotlib}. Training runs on NVIDIA GeForce GTX 1080 Ti and Quadro GP100 GPUs without compiler acceleration. Both GPUs belong to the Pascal generation, so the training setup does not favour any tagger through newer hardware or compiler speed-ups.

\begin{table}[htbp]
\centering
\caption{The seven jet taggers compared in this work, with the input each receives and its key architectural feature. The two additional ResNeXt training seeds of Section~\ref{sec:results:uncertainty} use $40\times40$ images.}
\label{tab:models}
\small
\setlength{\tabcolsep}{5pt}
\begin{tabular}{lll}
\toprule
Model & Input & Architecture \\
\midrule
BDT       & 19 substructure variables       & Gradient-boosted trees (XGBoost) \\
CNN       & Jet image, $40\times40$          & 2D convolutional network \\
ResNeXt   & Jet image, $64\times64$, 3 ch.   & ResNeXt-50 ($32\times4d$) \\
ParticleNet & Particle cloud, 7 features     & EdgeConv, static $k$NN graph ($k=16$), 3 blocks \\
ParT      & Particle cloud, 17 + pairwise    & Transformer, 4 particle + 2 class-attention blocks \\
UParT     & Particle cloud, 17 + pairwise    & As ParT, mean pooling, no class attention \\
GloParT   & Particle cloud, 17 + pairwise    & As ParT, different layer-norm placement \\
\bottomrule
\end{tabular}
\end{table}

\begin{table}[htbp]
\centering
\small
\caption{Training settings of the seven taggers. The neural networks use the Adam \cite{Adam} or AdamW \cite{AdamW} optimizer. All are trained from scratch on the same JetClass sample, keeping the checkpoint with the lowest validation loss.}
\label{tab:training}
\begin{tabular}{lcccc}
\toprule
Tagger & Optimiser & Learning rate & Batch size & Epochs \\
\midrule
BDT         & XGBoost & 0.05      & --   & $\leq2000$ rounds$^{*}$ \\
CNN         & Adam    & $10^{-4}$ & 512  & 30 \\
ResNeXt     & AdamW   & $10^{-4}$ & 256  & 20 \\
ParticleNet & Adam    & $10^{-3}$ & 1024 & $\leq20^{*}$ \\
ParT        & Adam    & $10^{-3}$ & 256  & 20 \\
UParT       & Adam    & $10^{-3}$ & 256  & 20 \\
GloParT     & Adam    & $10^{-3}$ & 256  & 20 \\
\bottomrule
\multicolumn{5}{l}{\footnotesize $^{*}$ with early stopping on the validation loss.}
\end{tabular}
\end{table}

\section{Results}
\label{sec:results}

\subsection{The Benchmark Ranking}
\label{sec:results:jetclass}

\begin{figure}[htbp]
\centering
\includegraphics[width=0.6\textwidth]{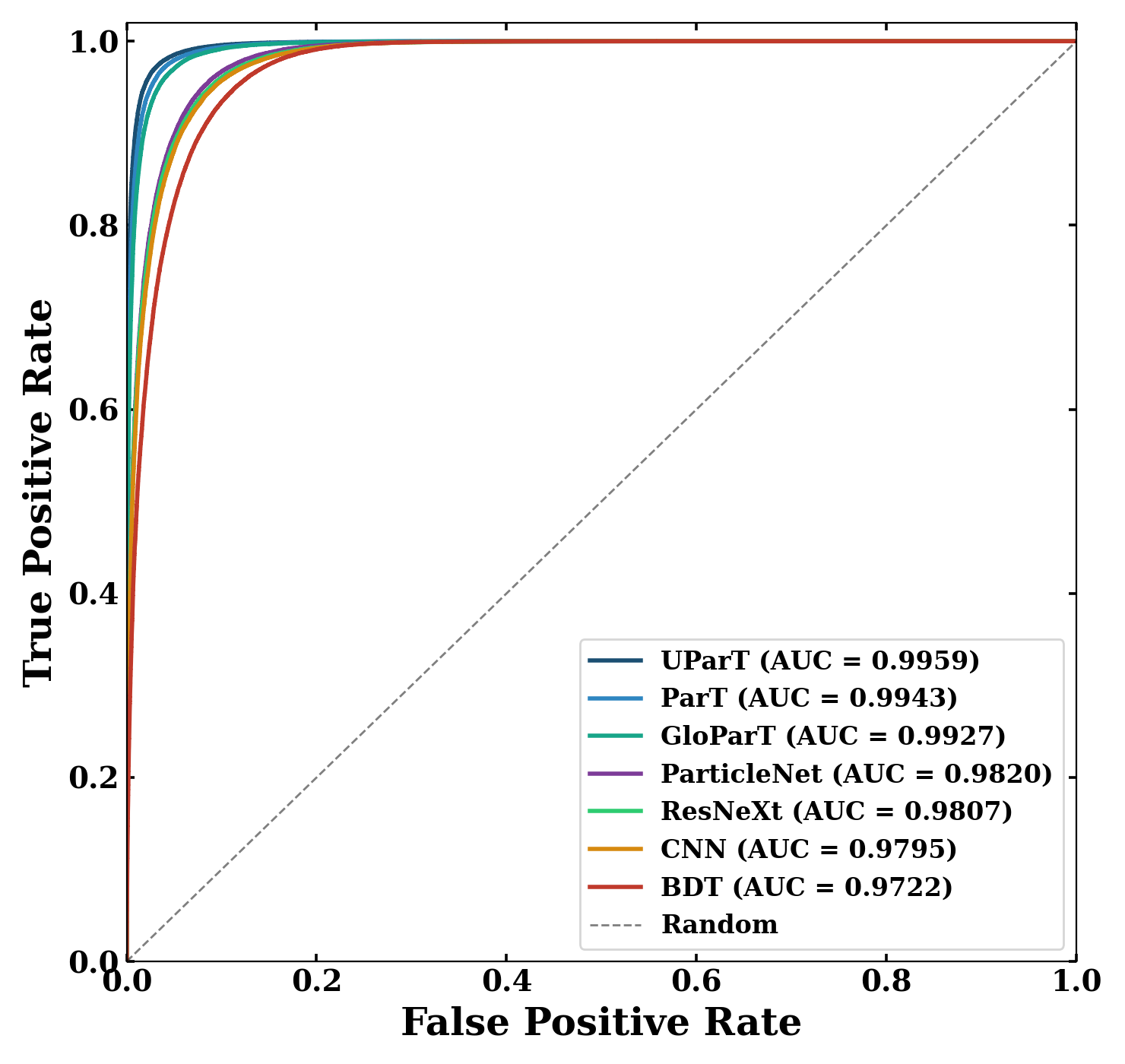}
\caption{Performance on the JetClass held-out test set at training
time, in the fixed $500$--$1000$~GeV training window and with the
split of Section~\ref{sec:training}. This is distinct from the
per-mass $Z'$ application sample used by every other ROC and AUC
figure in this paper. The curves show the true-positive rate (signal efficiency) against the false-positive rate (background efficiency), with each tagger's AUC in the legend.}
\label{fig:jetclass-training-roc-pair}
\end{figure}

The JetClass test set gives the reference ranking of the seven taggers within the $500$--$1000$~GeV jet-$p_T$ range used for training. Figure~\ref{fig:jetclass-training-roc-pair} shows this benchmark, which serves as the starting point for the search-level comparisons that follow. The AUC values split the taggers into three clear tiers. The three transformers lead with UParT at 0.9959, ParT at 0.9943, and GloParT at 0.9927. ParticleNet (0.9820) and the two image-based models (0.9807 and 0.9795) follow, and the BDT has the lowest AUC (0.9722). The rest of this paper tests whether this order and the size of these gaps survive in a realistic resonance search. In such a search, jets may lie outside the training $p_T$ range, detector effects and pileup are present, and the working point is set by its effect on discovery sensitivity.

\subsection{Discrimination In The $Z'$ Search And Its Mass Dependence}
\label{sec:results:zprime_disc}

In the $Z'$ search the benchmark ranking still predicts the order of the taggers, but it does not fully capture how their performance affects search sensitivity. The advantage of the transformers grows with resonance mass, and a single AUC value per tagger cannot describe this trend. Table~\ref{tab:auc} first checks whether the benchmark ranking holds on the application samples. The AUC is computed separately at each of the six $Z'$ mass points, and the table gives the unweighted mean and standard deviation of these six values for each tagger. The mean can be compared directly with the JetClass benchmark AUC, while the standard deviation shows how much the discrimination varies across the resonance-mass range.

\begin{table}[htbp]
\centering
\caption{AUC on the $Z'$ application sample, as the unweighted mean
over the six mass points ($1200$--$2500$~GeV). ``Spread'' is the
standard deviation of the six per-mass values and not a statistical
uncertainty. The statistical uncertainty on each per-mass AUC is at
most $\pm0.001$ (Section~\ref{sec:results:uncertainty}), much smaller than the
spread, which therefore reflects a genuine dependence on the resonance mass.}
\label{tab:auc}
\begin{tabular}{lcc}
\toprule
Model & Mean AUC (over 6 masses) & Spread across mass \\
\midrule
GloParT            & 0.928 & $\pm0.007$ \\
ParT               & 0.927 & $\pm0.010$ \\
UParT              & 0.911 & $\pm0.010$ \\
ParticleNet        & 0.897 & $\pm0.006$ \\
ResNeXt            & 0.896 & $\pm0.008$ \\
CNN                & 0.896 & $\pm0.010$ \\
BDT / XGBoost      & 0.887 & $\pm0.011$ \\
\bottomrule
\end{tabular}
\end{table}

Every tagger loses discrimination in the $Z'$ search, and its mean AUC is $0.065$--$0.085$ lower than its benchmark value in Section~\ref{sec:results:jetclass}. This loss is expected because the application samples add several challenges at once. The selected signal jets need not contain all the top-quark decay products, full detector simulation and pileup affect the inputs, and many jets lie above the training $p_T$ range. The two sets of AUC values therefore describe different things. The JetClass values measure discrimination under training conditions, while the $Z'$ values measure how much of that discrimination survives in the jets selected by the search. The $Z'$ values are the more direct measure for the search studied here. At the level of tagger groups the benchmark ranking is largely preserved. The transformers remain the leading group, although their internal order changes. GloParT and ParT lead in the search, while UParT, the best transformer on JetClass, falls about $0.016$--$0.017$ behind them. ParticleNet, ResNeXt and the CNN form a tight middle group, so ParticleNet loses the small lead over the image-based taggers that it held on JetClass. The BDT remains the weakest model. The benchmark thus gets the overall order right but understates the gap between the leading transformers and the BDT, which grows from $0.024$ on JetClass to $0.041$ in the search. The following results show that the transformer advantage increases further with higher resonance masses.

\begin{figure}[htbp]
\centering
\includegraphics[width=\textwidth]{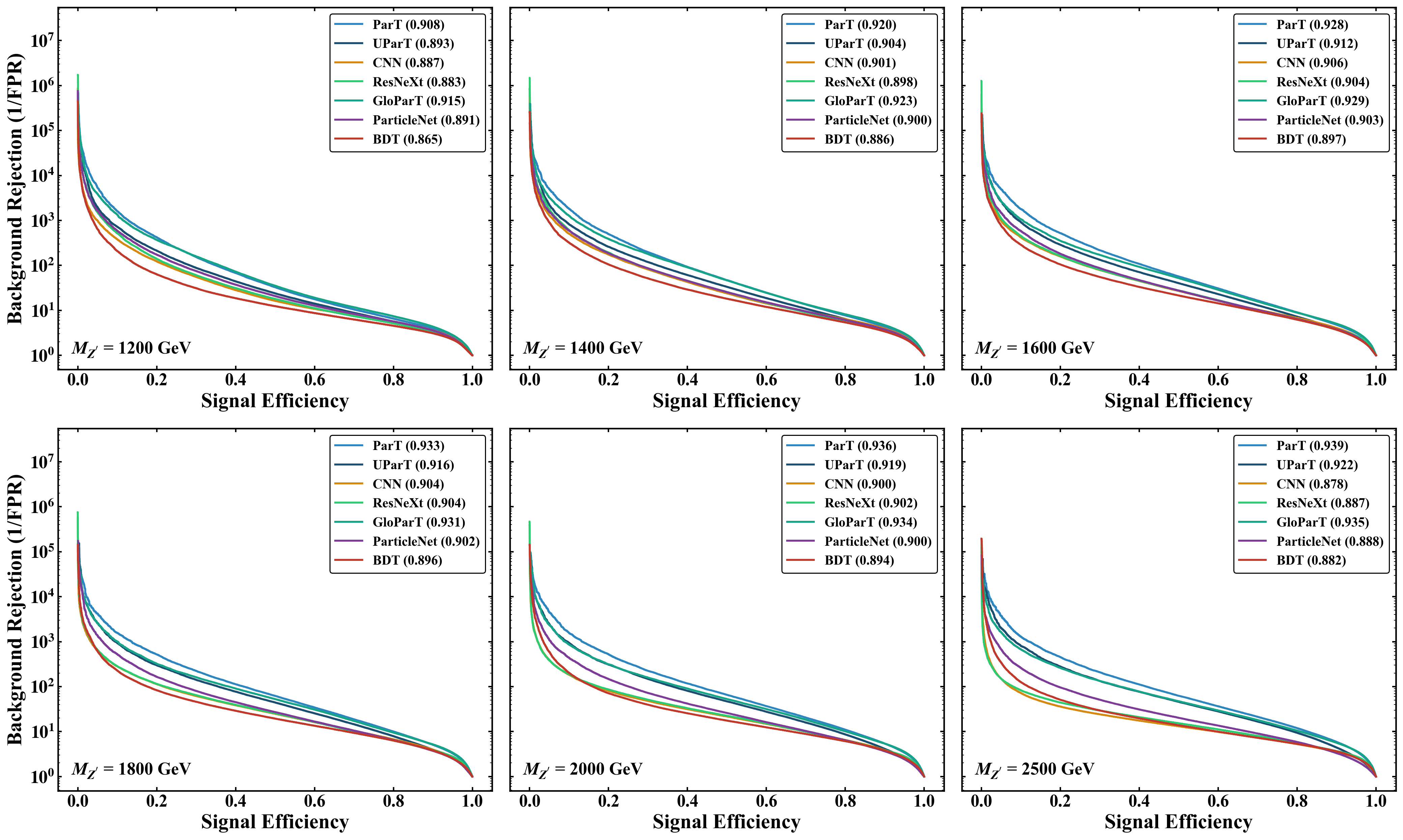}
\caption{Background rejection vs.\ signal efficiency for all seven
taggers, one panel per $Z'$ mass point, with each tagger's AUC in the
legend.}
\label{fig:roc-by-mass}
\end{figure}

The spread of AUC across the six mass points is small and similar for all taggers, with standard deviations ranging from $0.006$ to $0.011$. The per-mass ROC curves in Figure~\ref{fig:roc-by-mass} nonetheless show clear differences in how the taggers respond to rising resonance mass. The AUC of the transformers improves steadily from $1200$ to $2500$~GeV. ParticleNet, the image-based taggers, and the BDT improve only up to about $1600$~GeV and then slowly decline. This pattern is consistent with the transformers making better use of constituent-level information as the top-quark decay products become more collimated at high $p_T$. The gain matters most at high resonance mass, where strong rejection of the QCD background is needed to extend the search's reach. The benchmark AUC, therefore, captures the overall order but misses the growing lead of transformers in the high-mass region.

\begin{figure}[htbp]
\centering
\includegraphics[width=0.65\textwidth]{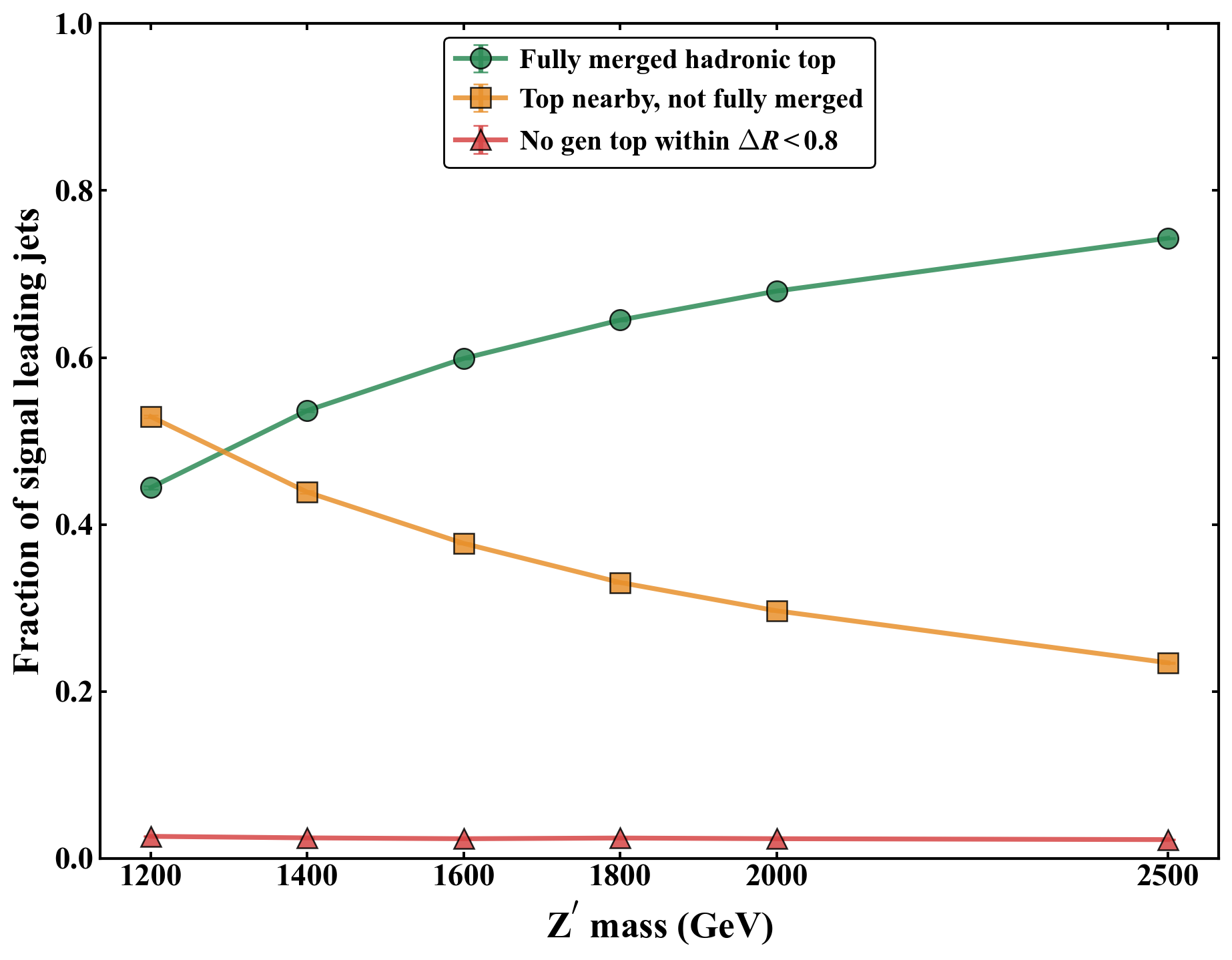}
\caption{Fraction of signal jets that are fully merged, partially
merged, or unmatched to a generator-level top, by $Z'$ mass. The
binomial statistical uncertainties are smaller than the markers.}
\label{fig:truth-fractions}
\end{figure}

Part of this mass dependence stems from changes in signal composition rather than from the taggers themselves. The leading jet is not required to contain all the top-quark decay products, and the share of jets that do contain them rises with the top-quark boost. To separate the two effects, each signal jet is matched to the nearest generator-level top quark and labeled fully merged, partially merged, or unmatched. A jet is fully merged if the $b$ quark and both quarks from the $W$-boson decay lie within $\Delta R<0.8$ of the jet axis. This is the same definition used to select top jets in JetClass. Fully merged jets are therefore the closest match to the signal jets seen in training. Partially merged jets form a different topology that the training selection does not cover, and they pull down the inclusive AUC. As the top quarks become more boosted, their decay products become more collimated. The fully merged fraction therefore rises from $44\%$ at $m_{Z'}=1200$~GeV to $74\%$ at $2500$~GeV (Figure~\ref{fig:truth-fractions}).

\begin{figure}[htbp]
\centering
\includegraphics[width=\textwidth]{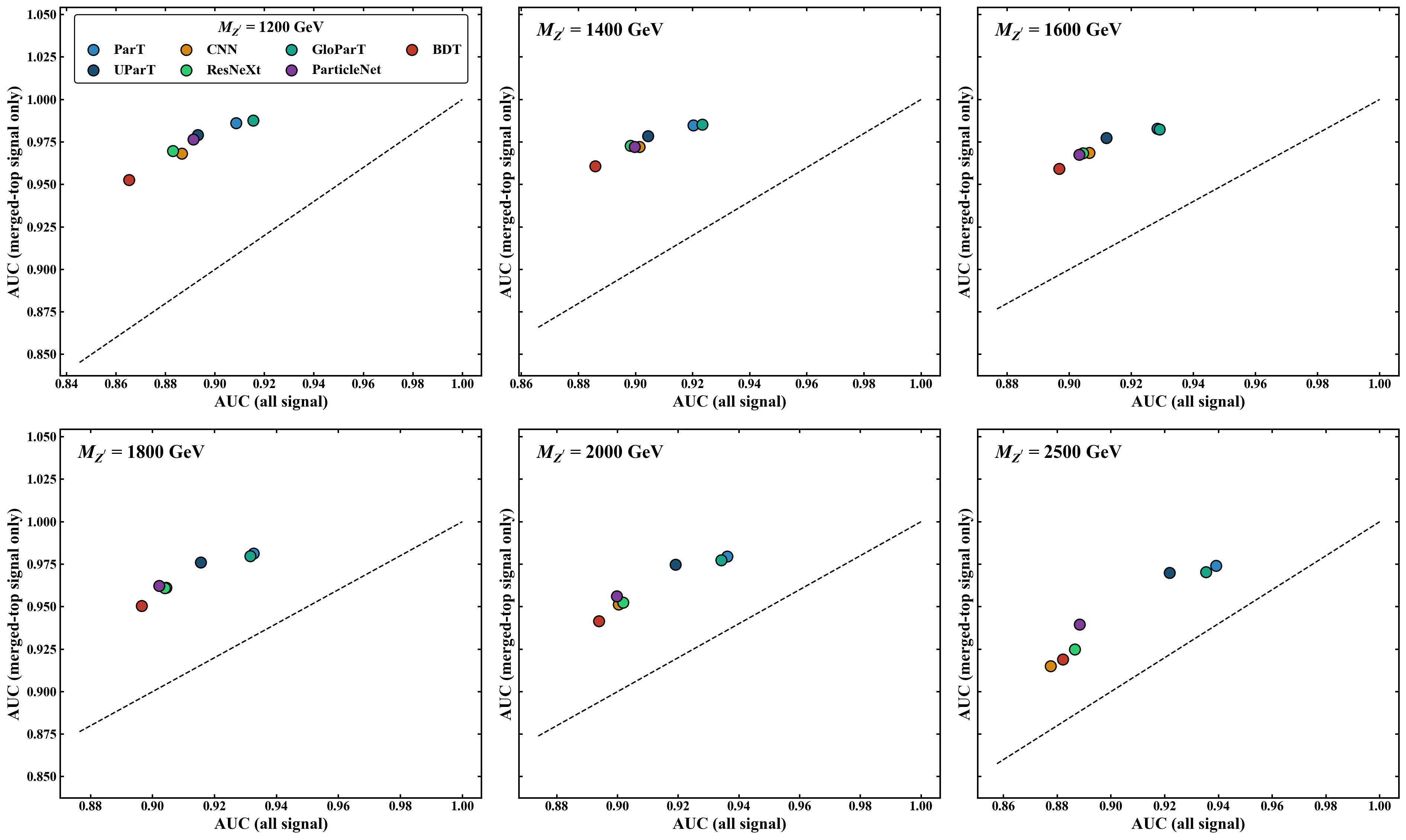}
\caption{AUC using all signal jets vs.\ AUC using only fully merged
signal jets, one panel per mass point. The diagonal marks equal AUC.}
\label{fig:truth-matching}
\end{figure}

Restricting the comparison to fully merged jets reverses the mass dependence, as Figure~\ref{fig:truth-matching} shows. The AUC values are much higher, between $0.953$ and $0.988$ at $1200$~GeV, but they fall with resonance mass for every tagger. The transformers lose about $0.01$--$0.02$ between the lowest and highest mass points, while the other taggers lose about $0.03$--$0.05$. Two effects therefore compete. The growing fraction of fully merged tops raises the AUC with mass, while each tagger becomes worse at recognizing a merged top as $p_T$ increases. The transformers lose least to the second effect and so pull ahead. The next section relates this advantage to jet $p_T$ above the training range. For a search, the choice of tagger therefore matters most where the signal is rarest. At high resonance mass, the cross section is smallest, and the gap between the transformers and the other taggers is largest.

\subsection{Extrapolation Beyond the Training Range}
\label{sec:results:domainshift}
\label{subsubsec:pt-mismatch}

The mass dependence above is closely tied to how the taggers extrapolate beyond their training $p_T$ range. A heavier resonance produces more energetic top quarks and therefore harder jets, many of which lie above the $500$--$1000$~GeV training range. The results indicate that the taggers differ not only in how well they identify top jets inside the training range but also in how well they keep this discrimination at higher $p_T$. The ranking in Section~\ref{sec:results:zprime_disc} is measured on the $Z'$ application samples, while all taggers are trained only on JetClass. Its value for the search therefore depends on how well performance carries over across the differences between the training and application samples listed in Table~\ref{tab:train_vs_apply}. This section and the next examine two parts of this mismatch. These are the extrapolation beyond the fixed training $p_T$ range and the effect of pileup. To separate the dependence on jet $p_T$ from the dependence on resonance mass, each tagger is evaluated in bins of reconstructed jet $p_T$ with edges at $400$, $500$, $600$, $750$, $850$, $1000$, $1300$ and $1600$~GeV. At each resonance mass, only the bins inside the corresponding $p_T$ selection window are used. The lowest bin lies below the JetClass training range, and the two highest bins lie above it.

\begin{figure}[htbp]
\centering
\includegraphics[width=\textwidth]{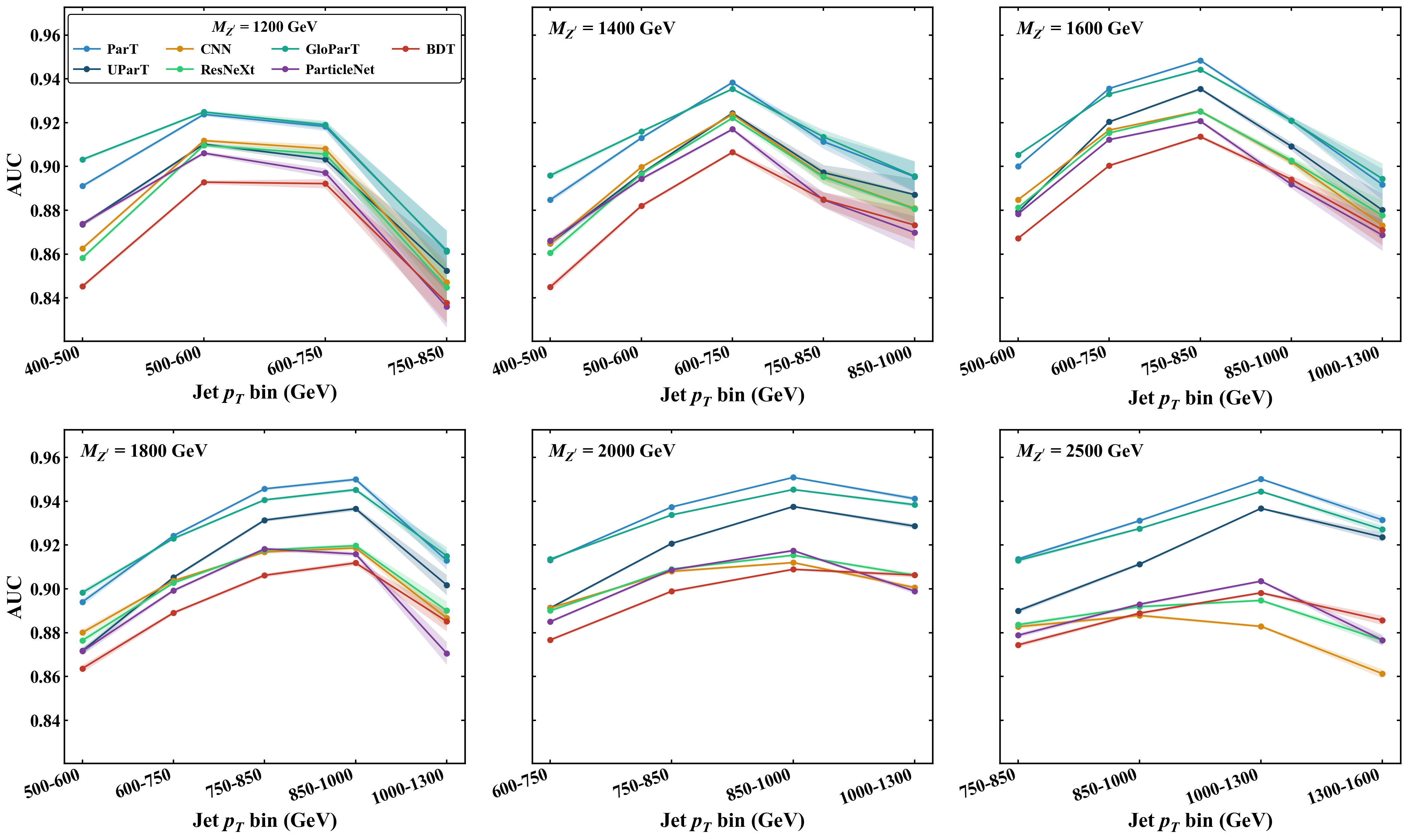}
\caption{AUC by jet $p_T$ bin, one panel per $Z'$ mass point, for all
seven taggers. Each panel shows the bins inside that mass point's
window, and the shaded bands are the statistical uncertainties discussed in Section~\ref{sec:results:uncertainty}.}
\label{fig:auc-ptbin-by-mass}
\end{figure}

\begin{figure}[htbp]
\centering
\includegraphics[width=\textwidth]{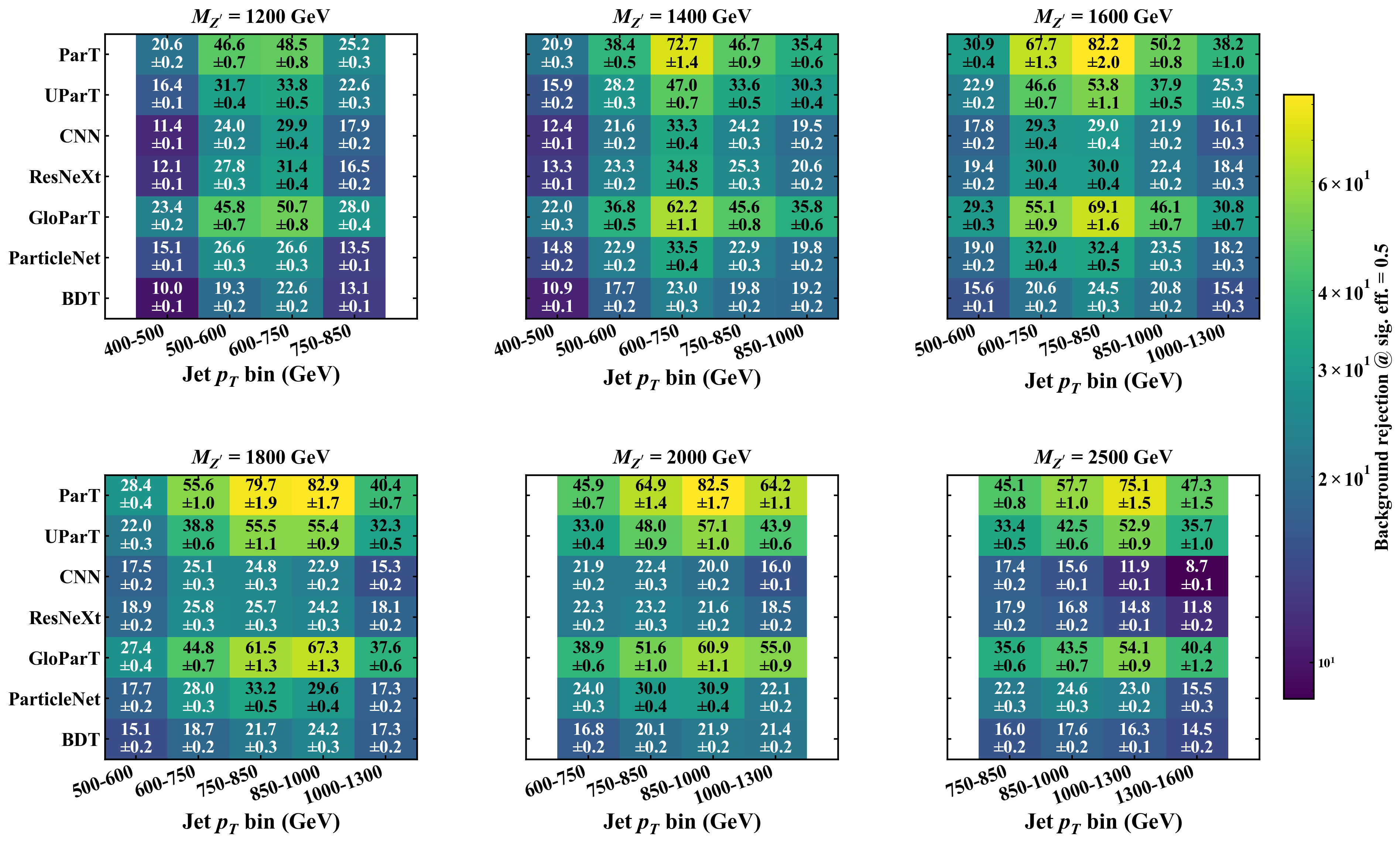}
\caption{Background rejection at $\varepsilon_{\rm sig}=0.5$, by jet
$p_T$ bin and $Z'$ mass point, for all seven taggers.}
\label{fig:heatmap-bkgrej-ptbin-by-mass}
\end{figure}

\begin{figure}[htbp]
\centering
\includegraphics[width=\textwidth]{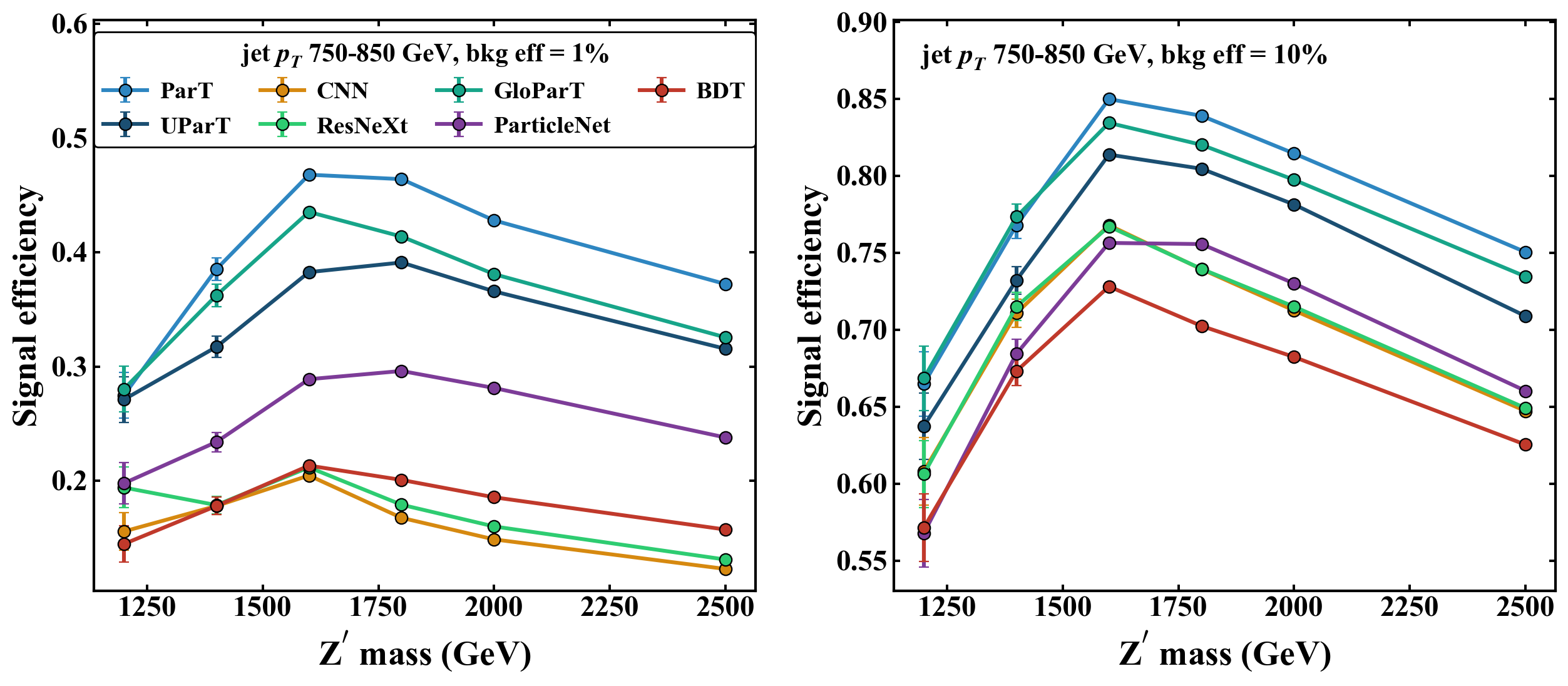}
\caption{Signal efficiency vs.\ $Z'$ mass for all seven taggers at
fixed background efficiency of $1\%$ (left) and $10\%$ (right), for
jets with $750 < p_T < 850$~GeV.}
\label{fig:pnet-signal-eff}
\end{figure}

Figure~\ref{fig:auc-ptbin-by-mass} shows the AUC in each bin, and Figure~\ref{fig:heatmap-bkgrej-ptbin-by-mass} shows the background rejection at a signal efficiency of $0.5$ so that the taggers are compared at a common signal efficiency. Below the training range, all taggers lose discrimination, but the transformers and ParticleNet are less affected than the BDT and the image-based models. The same difference appears above the training range, where the transformers hold their performance better as jet $p_T$ increases. At $m_{Z'}=2500$~GeV, for example, the AUC of ParT rises from $0.931$ in the $850$--$1000$~GeV bin to $0.950$ in the $1000$--$1300$~GeV bin, while the background rejection of the other taggers falls beyond $850$~GeV. Both figures therefore show that the transformers keep their discrimination best when extrapolating above the training $p_T$ range. A further test checks whether jet $p_T$ alone drives the mass dependence. For this test, the taggers are compared in the common $750$--$850$~GeV bin across all six resonance masses (Figure~\ref{fig:pnet-signal-eff}). The signal efficiency at fixed background efficiency still changes with resonance mass for every tagger, so jet $p_T$ alone does not explain the behaviour. The transformers nonetheless maintain the highest signal efficiency across the entire mass range. At $1\%$ background efficiency, ParT retains $37$--$47\%$ of the signal compared with $12$--$21\%$ for the CNN. The transformer advantage is therefore not limited to jets beyond the training range and also holds for jets within a common $p_T$ interval.

These results link the $p_T$-dependent performance to the mass dependence seen in Section~\ref{sec:results:zprime_disc}. As the resonance mass increases, a larger share of the signal lies above the $1000$~GeV training boundary, where the transformers maintain better discrimination than the other taggers. This exposes a limitation of benchmarking in a restricted $p_T$ range. Performance measured inside that range need not represent a tagger across the full kinematic range of a search. In this study, the transformer advantage is largest in a region outside the range used for the benchmark ranking. The $p_T$-binned comparison does not, however, single out extrapolation as the only cause. Other properties also change with jet $p_T$, such as the fraction of fully merged top jets and the detector response. The BDT and image-based taggers also receive fewer constituents in the application samples than in training (Section~\ref{sec:reco}). The observed differences may therefore come from a mix of these effects, and the present study cannot fully separate their individual contributions. Training and evaluating the taggers over a wider $p_T$ range would directly test their extrapolation.

\subsection{Pileup}
\label{subsubsec:pileup}

Robustness to pileup is a common concern for taggers trained on benchmark datasets, because pileup is absent from JetClass but present in real collision data. Its effect is assessed using the AUC as a function of the number of reconstructed primary vertices ($n_{\mathrm{PV}}$) across all six resonance masses, with the full background sample. Pileup also adds energy to the reconstructed jets. At $m_{Z'}=1200$~GeV, the mean signal-jet $p_T$ rises by about $20$~GeV from the lowest to the highest $n_{\mathrm{PV}}$ bin. Since the AUC depends strongly on jet $p_T$ (Section~\ref{subsubsec:pt-mismatch}), each $n_{\mathrm{PV}}$ bin is reweighted to the inclusive jet-$p_T$ distribution, separately for signal and background. This keeps changes in the jet-$p_T$ spectrum from hiding the direct effect of pileup.

\begin{figure}[htbp]
\centering
\includegraphics[width=\textwidth]{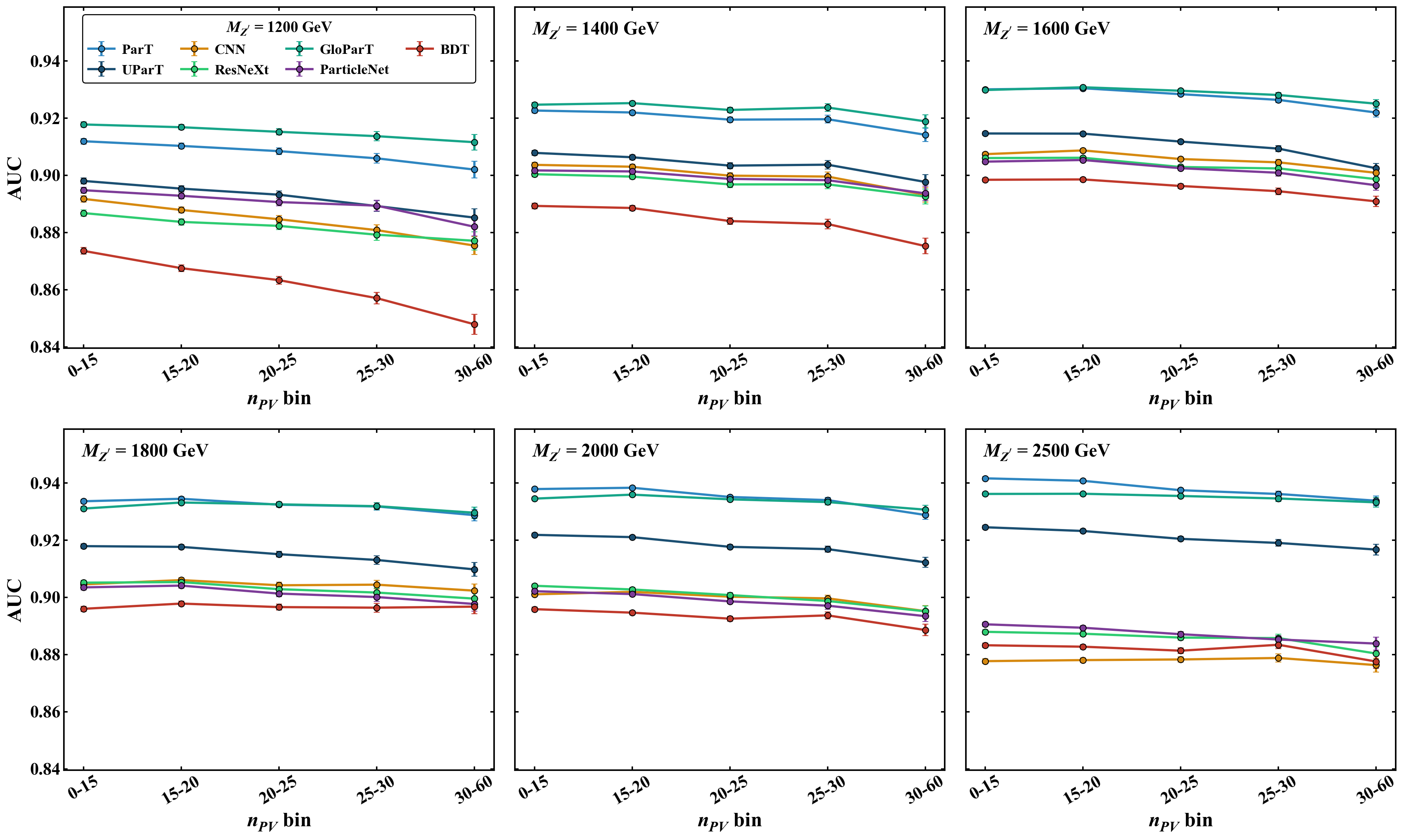}
\caption{AUC vs.\ $n_{\mathrm{PV}}$ bin at all six $Z'$ mass points,
one panel per mass, after reweighting the jet-$p_T$ spectrum of each
$n_{\mathrm{PV}}$ bin to the inclusive one.}
\label{fig:pileup-by-mass}
\end{figure}

\begin{figure}[htbp]
\centering
\includegraphics[width=0.7\textwidth]{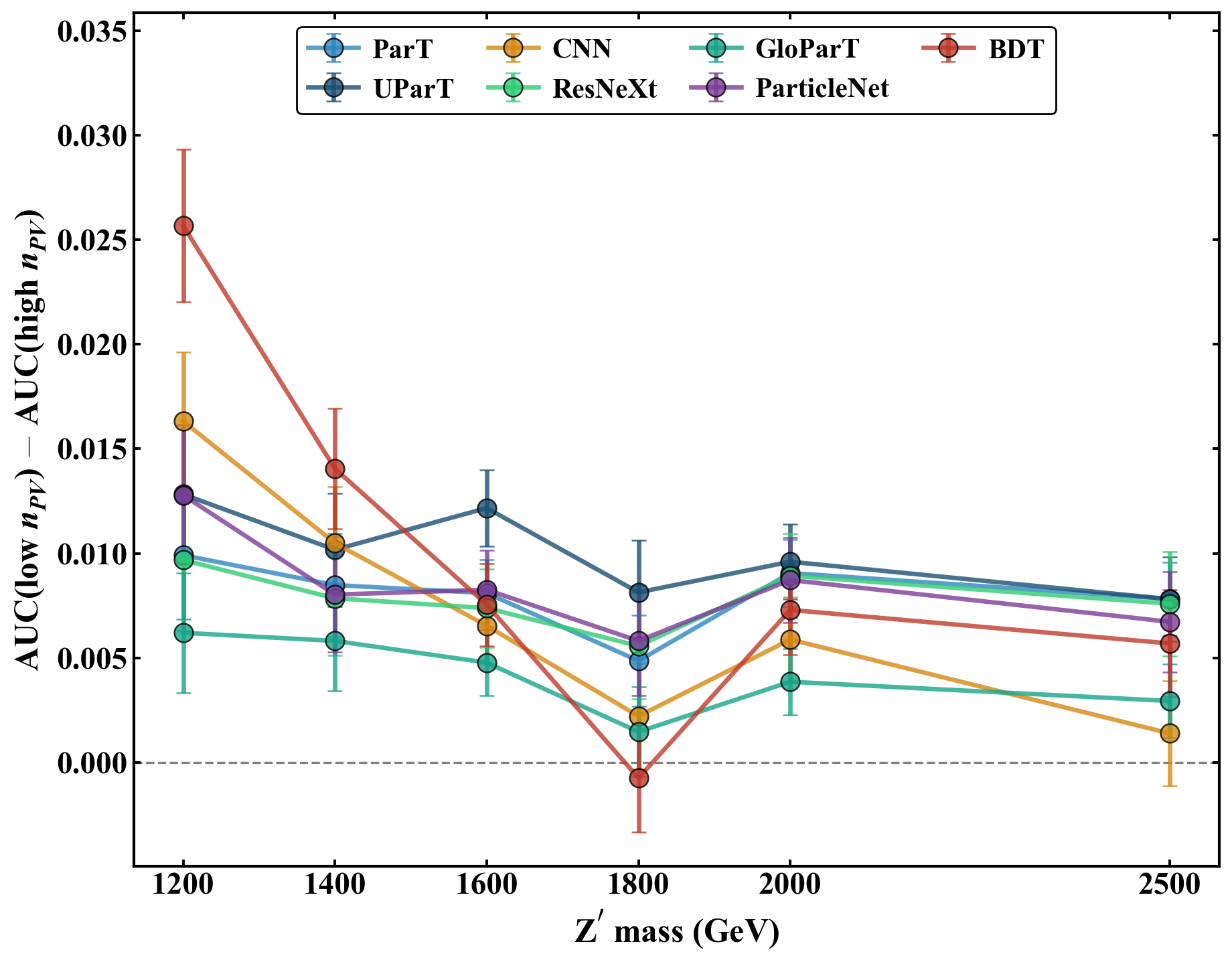}
\caption{AUC lost from the lowest ($0$--$15$) to the highest
($30$--$60$) $n_{\mathrm{PV}}$ bin, by tagger and $Z'$ mass, after
the jet-$p_T$ reweighting, with the statistical uncertainty on the
difference. The dashed line marks no change.}
\label{fig:pileup-degradation}
\end{figure}

Pileup has only a modest effect on the discrimination of the taggers, and the loss does not grow systematically with resonance mass (Figures~\ref{fig:pileup-by-mass} and~\ref{fig:pileup-degradation}). GloParT is among the most robust models, while the BDT degrades somewhat more at lower masses. The overall ranking of the taggers stays essentially unchanged, with only minor shifts among models of similar performance. This stability agrees with the robustness reported for ParT in quark--gluon tagging under the higher pileup expected at the HL-LHC \cite{Castillo:2025qg}. Under the 2016 pileup conditions used here, pileup therefore has a limited effect on the relative performance of the taggers even without pileup mitigation on the jet constituents. Within the scope of this study, pileup does not substantially change the choice of tagger for the search. The mismatch between the training and application $p_T$ ranges has a larger effect on how well benchmark performance carries over to the search.

\subsection{From Discrimination To Significance}

\label{sec:results:significance}

As noted in Section~\ref{sec:intro}, the outcome of a search depends on the signal and background yields that survive a cut on the tagger score in the high-rejection region of the ROC curve. Small differences in AUC can therefore turn into much larger differences in discovery sensitivity \cite{Kasieczka:2015jma}. The reason is clear in the background-dominated regime ($S \ll B$), where the Asimov significance approaches $Z \simeq S/\sqrt{B}$. Stronger background rejection can then substantially increase the significance, as long as enough signal survives. A more powerful tagger can sustain tighter thresholds while keeping a useful signal yield, whereas a weaker tagger loses signal faster as the selection tightens.

The taggers are compared through the Asimov significance
\cite{CowanAsimov},
\begin{equation}
\label{eq:asimov_significance}
Z = \sqrt{2\left[(S+B)\ln\left(1+\frac{S}{B}\right)-S\right]}
\end{equation}

where $S$ and $B$ are the weighted signal and background yields that pass the tagger selection. This quantity is called the jet-level significance because the selection is applied to the leading jet within the mass-dependent $p_T$ window. A full resonance search~\cite{CMSZprimeBenchmark} would instead tag both top-quark jets and fit the reconstructed $t\bar{t}$ invariant-mass spectrum. For independently tagged jets, the signal and background efficiencies then scale roughly as the square, so the jet-level differences between taggers found here are expected to grow in such a search. The significance and mass reach in this paper are therefore intended to compare the taggers, not to predict the sensitivity of a complete search. The signal and background yields are normalized with leading-order cross sections from MadGraph and Pythia~8 after matching and filtering with the CMSSW \texttt{GenXSecAnalyzer}. The signal cross section falls from $1745.0$~fb at $m_{Z'}=1200$~GeV to $47.11$~fb at $2500$~GeV. The background cross sections are $3.80\times10^{7}$~fb for QCD and $3.78\times10^{5}$~fb for $t\bar{t}$. Scale and PDF uncertainties are not included. Two working-point strategies are used. The first fixes the signal efficiency at $50\%$ and gives a common reference for all taggers. The second finds the optimal working point separately for each tagger and mass point. It scans the signal efficiency in steps of $0.01$ and picks the threshold that maximizes $Z$. Each candidate threshold must keep at least 10 simulated background jets to limit statistical fluctuations. The optimization uses even-numbered events and is evaluated on odd-numbered events; the roles are then swapped. The final significance comes from the combined weighted yields of the two evaluation samples.

\begin{figure}[htbp]
\centering
\begin{subfigure}[b]{0.48\textwidth}
\centering
\includegraphics[width=\textwidth]{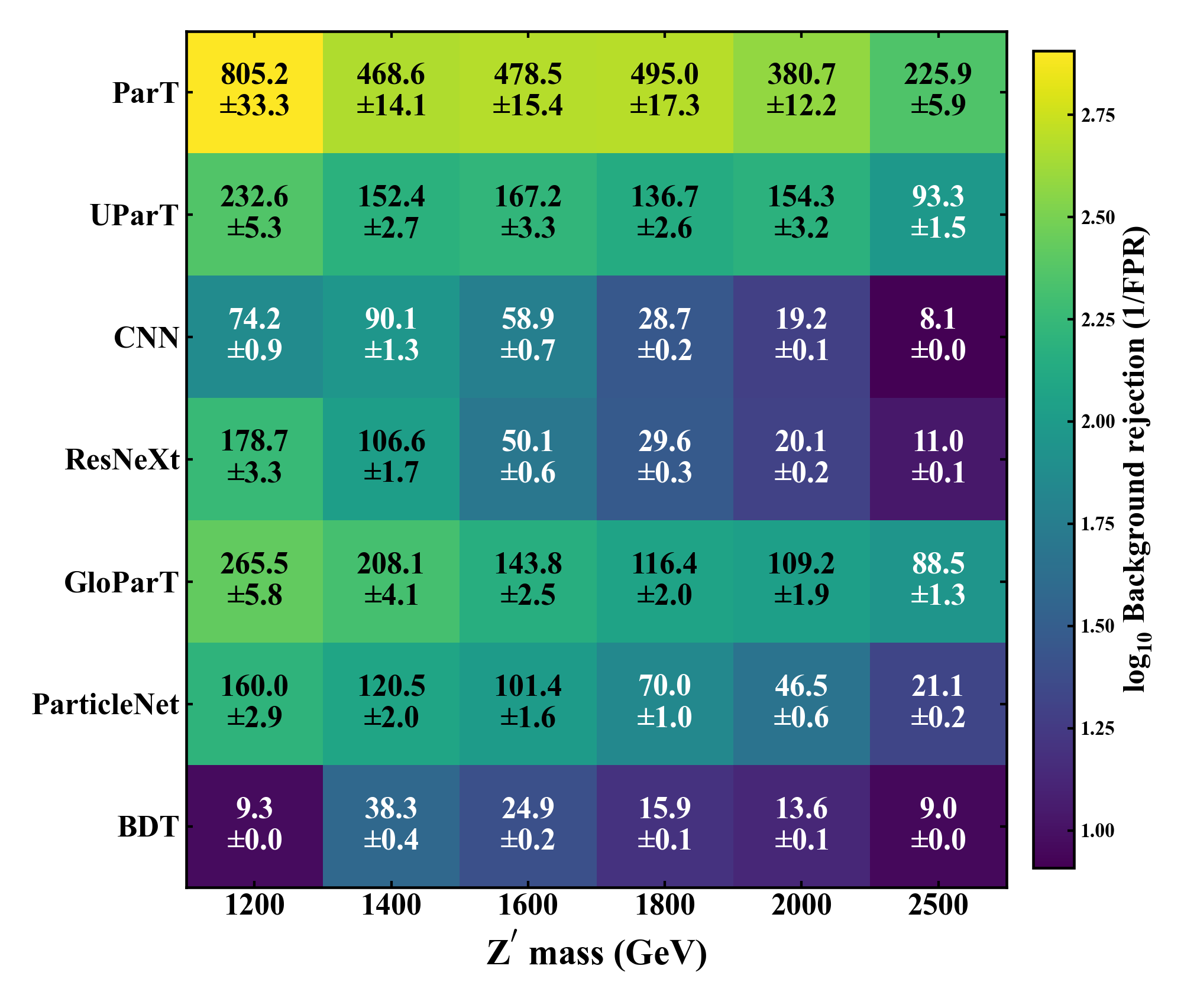}
\caption{}
\label{fig:heatmap-bkgrej-optimal}
\end{subfigure}
\hfill
\begin{subfigure}[b]{0.48\textwidth}
\centering
\includegraphics[width=\textwidth]{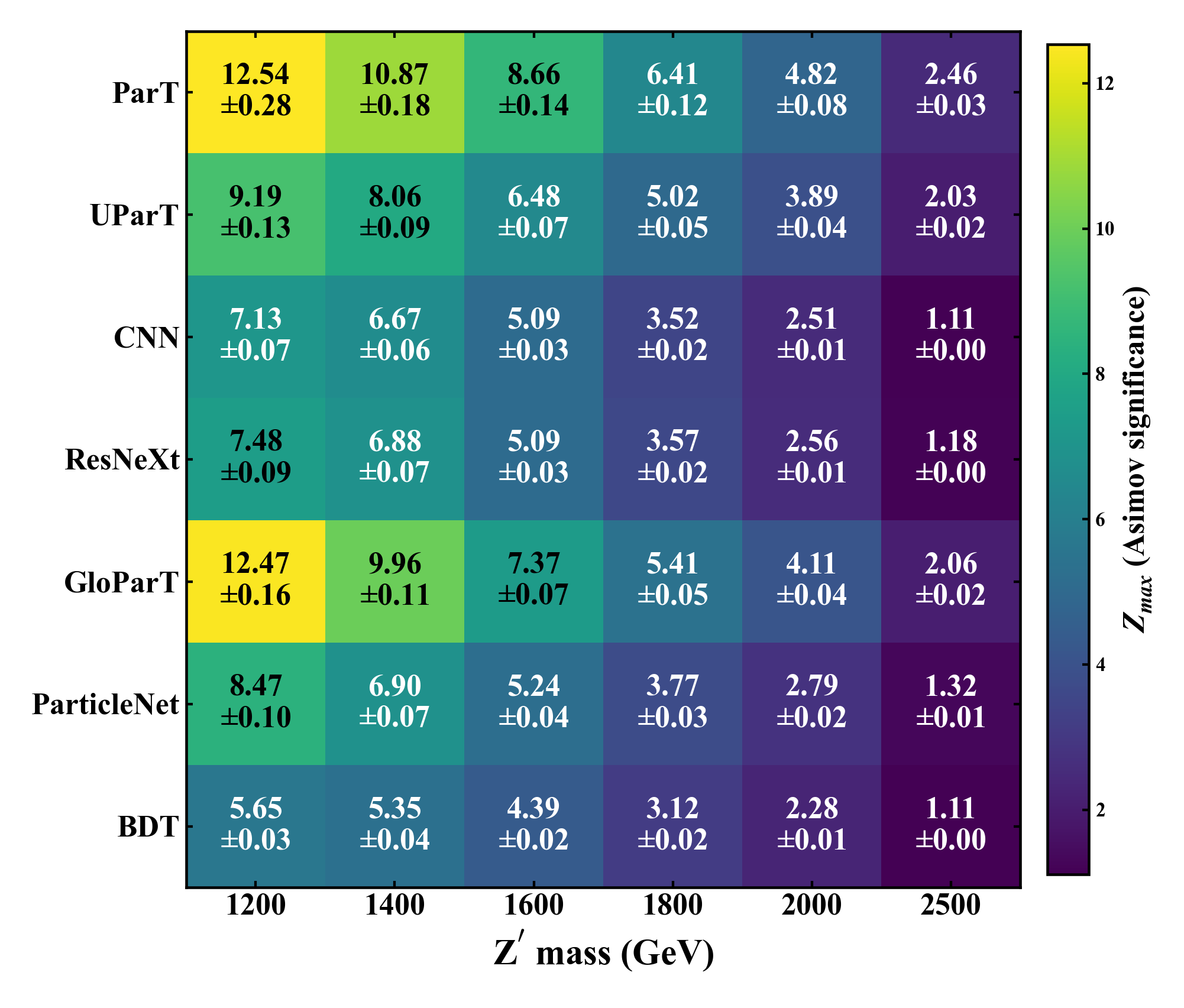}
\caption{}
\label{fig:heatmap-Z-optimal}
\end{subfigure}
\caption{Per-mass-optimised working point by tagger and $Z'$ mass
with statistical uncertainties. \textbf{(a)} Background rejection at
that cut. \textbf{(b)} Asimov significance $Z$ at the same cut.}
\label{fig:optimal-wp-heatmaps}
\end{figure}

\begin{figure}[htbp]
\centering
\begin{subfigure}[b]{0.48\textwidth}
\centering
\includegraphics[width=\textwidth]{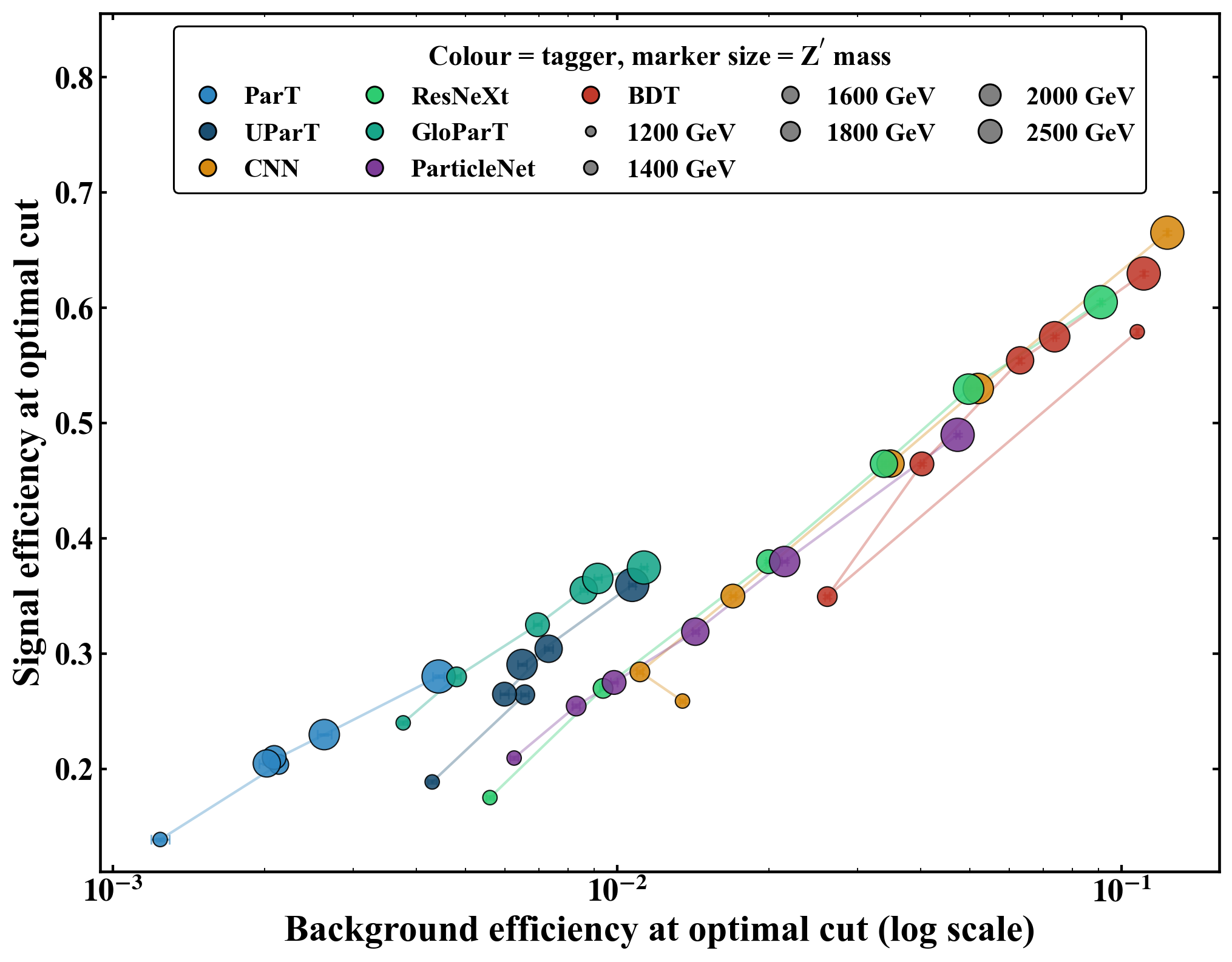}
\caption{}
\label{fig:efficiency-scatter}
\end{subfigure}
\hfill
\begin{subfigure}[b]{0.48\textwidth}
\centering
\includegraphics[width=\textwidth]{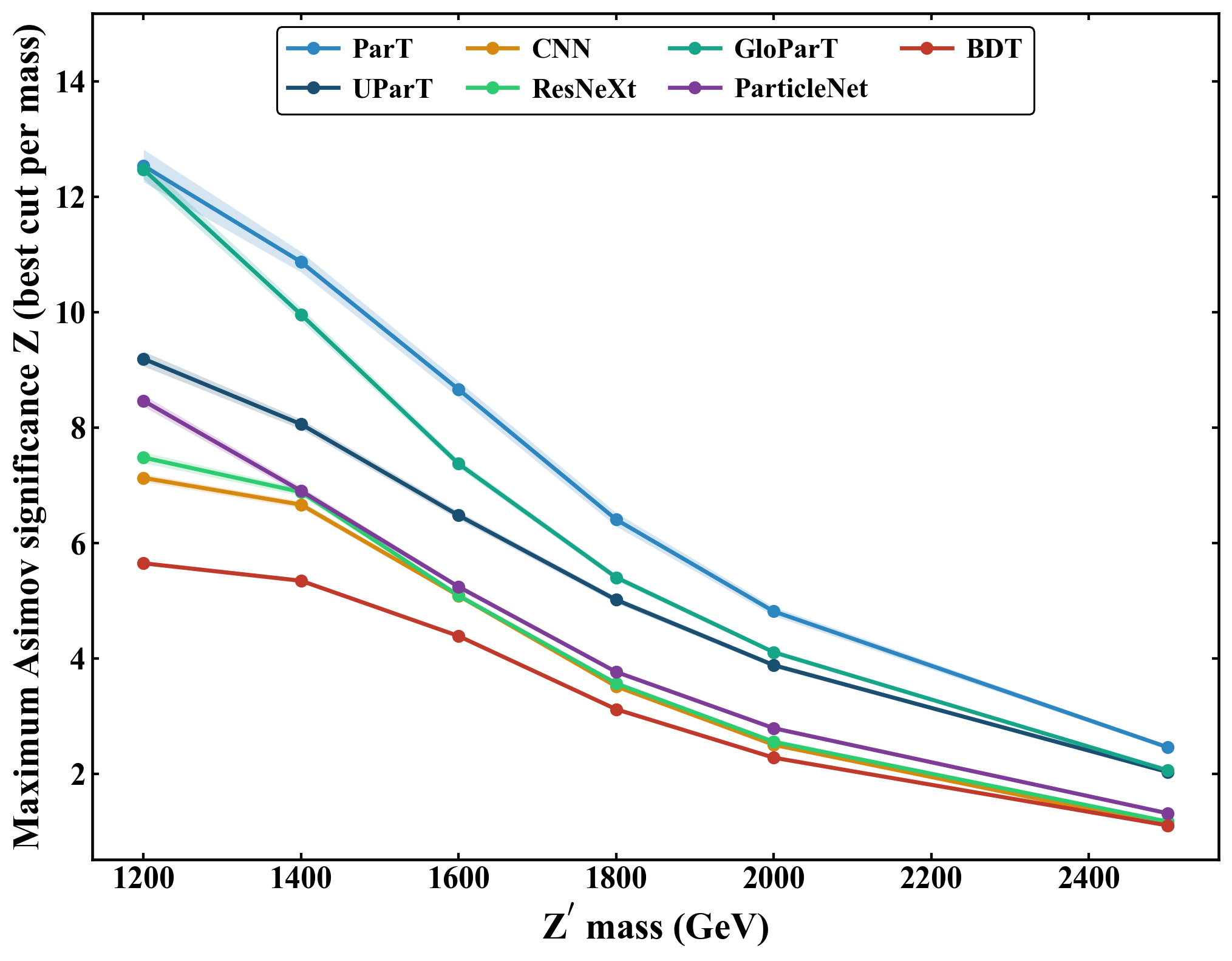}
\caption{}
\label{fig:zsig-vs-mass-optimal-plain}
\end{subfigure}
\caption{Per-mass-optimised working point shown as markers and lines.
\textbf{(a)} Signal vs.\ background efficiency at that cut, coloured
by tagger and sized by $Z'$ mass. \textbf{(b)} Asimov significance
$Z$ vs.\ $Z'$ mass with the values of Figure~\ref{fig:heatmap-Z-optimal}
drawn as one line per tagger.}
\label{fig:optimal-wp-lines}
\end{figure}

The optimized working points show clear differences in the background rejection of the taggers (Figures~\ref{fig:optimal-wp-heatmaps} and~\ref{fig:optimal-wp-lines}). ParT reaches the strongest rejection, closely followed by GloParT and UParT. ParticleNet and the image-based taggers sit in the middle, and the BDT gives the weakest rejection. The optimal significance falls with resonance mass for all taggers, mainly because the signal cross section drops. ParT and GloParT reach similar significance at $1200$~GeV, and ParT gives the highest significance at every higher mass point.

\begin{figure}[htbp]
\centering
\includegraphics[width=\textwidth]{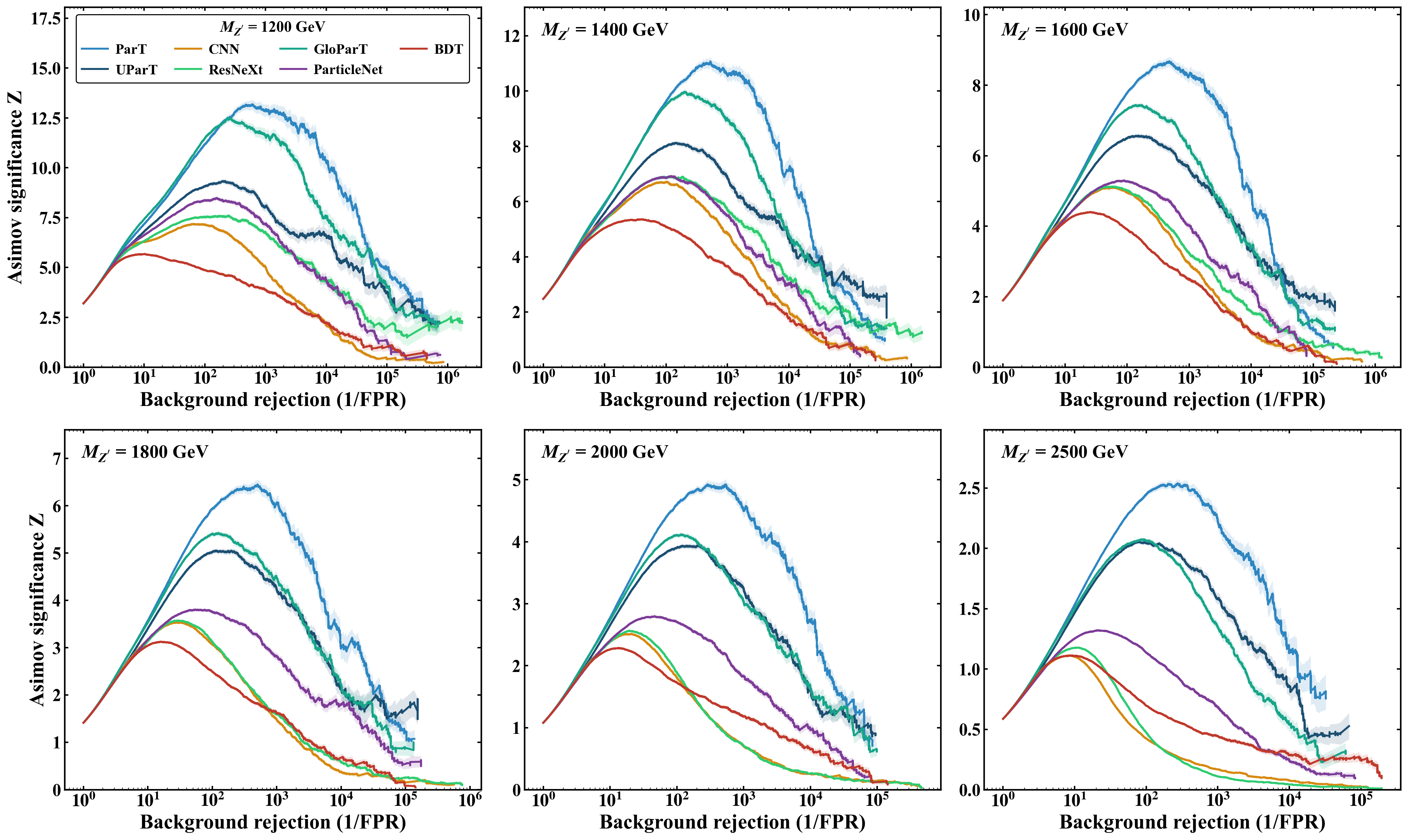}
\caption{Asimov significance $Z$ vs.\ background rejection for all
seven taggers with one panel per $Z'$ mass point. The maximum of each
curve is the optimal working point of Figure~\ref{fig:heatmap-Z-optimal}.}
\label{fig:zsig-vs-bkgrej-by-mass}
\end{figure}

\begin{figure}[htbp]
\centering
\includegraphics[width=\textwidth]{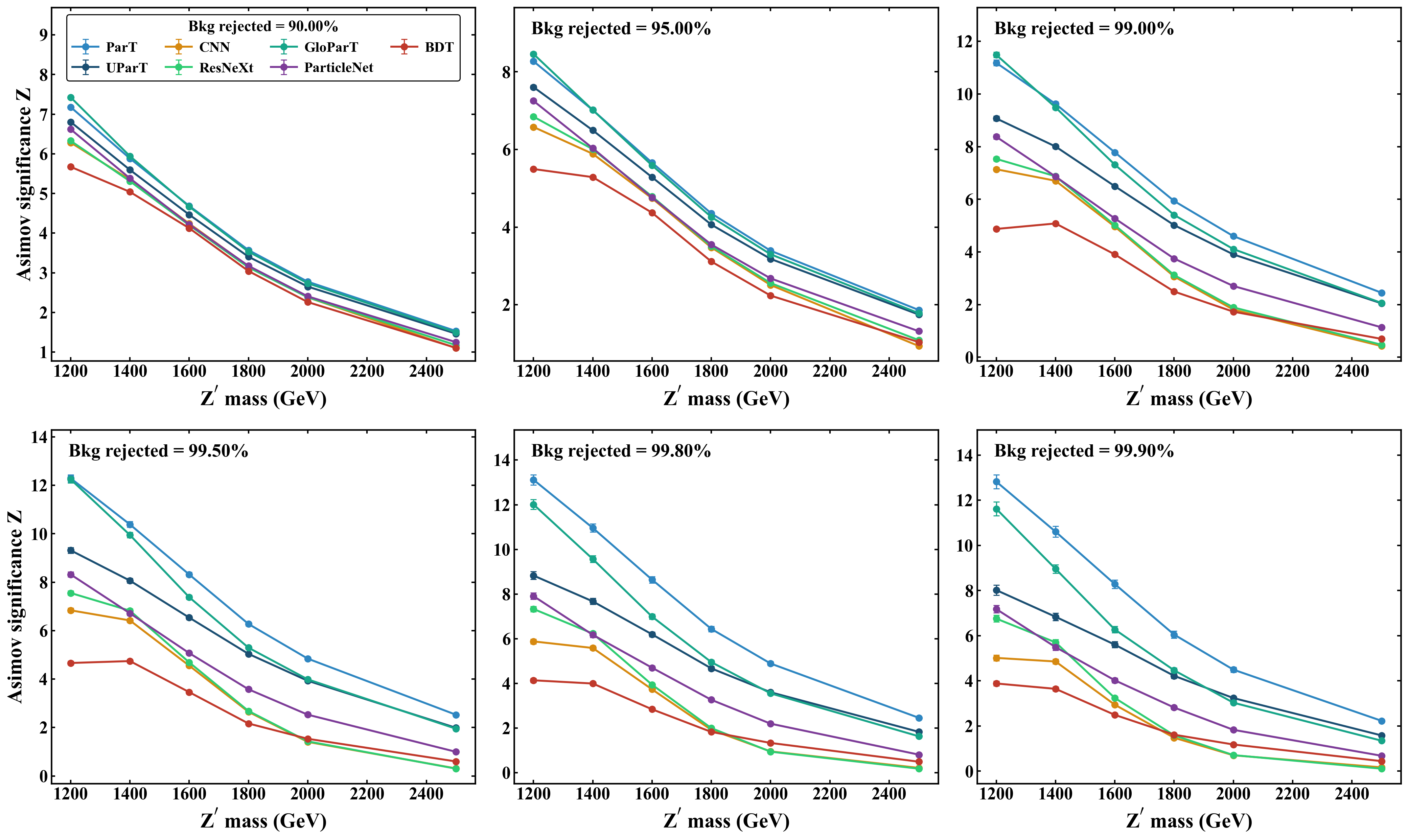}
\caption{Asimov significance $Z$ vs.\ $Z'$ mass at six fixed
background-rejection working points (90\%, 95\%, 99\%, 99.5\%, 99.8\%
and 99.9\%).}
\label{fig:zsig-vs-mass-by-rejection}
\end{figure}

Figure~\ref{fig:zsig-vs-bkgrej-by-mass} shows the significance as a function of background rejection for each tagger. Every tagger has an optimal working point that balances background suppression against signal efficiency. ParT reaches its maximum significance at much higher background rejection than the BDT, which gives the lowest significance across the mass range. Figure~\ref{fig:zsig-vs-mass-by-rejection} shows how the gap widens at tighter working points. The ratio of the best transformer to the BDT grows from $1.3$ at $90\%$ background rejection to $3.3$ at $99.9\%$ rejection. Differences in the high-rejection region can therefore be far larger than the AUC values alone suggest. The same pattern appears in the jet-level discovery reach. At the optimal working points, the transformers keep $Z\geq5$ up to $1800$~GeV, compared with $1600$~GeV for ParticleNet and the image-based taggers and $1400$~GeV for the BDT. These values show the relative sensitivity of the taggers under the jet-level definition used here and are not the projected reach of a complete resonance search. The fixed working point at $50\%$ signal efficiency gives a complementary view. The overall ranking stays broadly the same, with GloParT highest at $1200$~GeV and ParT leading at all higher masses. The transformers also keep a smaller fraction of the background than the other taggers. Their advantage at both fixed and optimized working points shows that transformer taggers can improve the sensitivity of a $Z'\to t\bar{t}$ search, especially in the high-rejection region.

\subsection{Inference Cost}
\label{sec:results:compute}

\begin{figure}[htbp]
\centering
\begin{subfigure}[b]{0.48\textwidth}
\centering
\includegraphics[width=\textwidth]{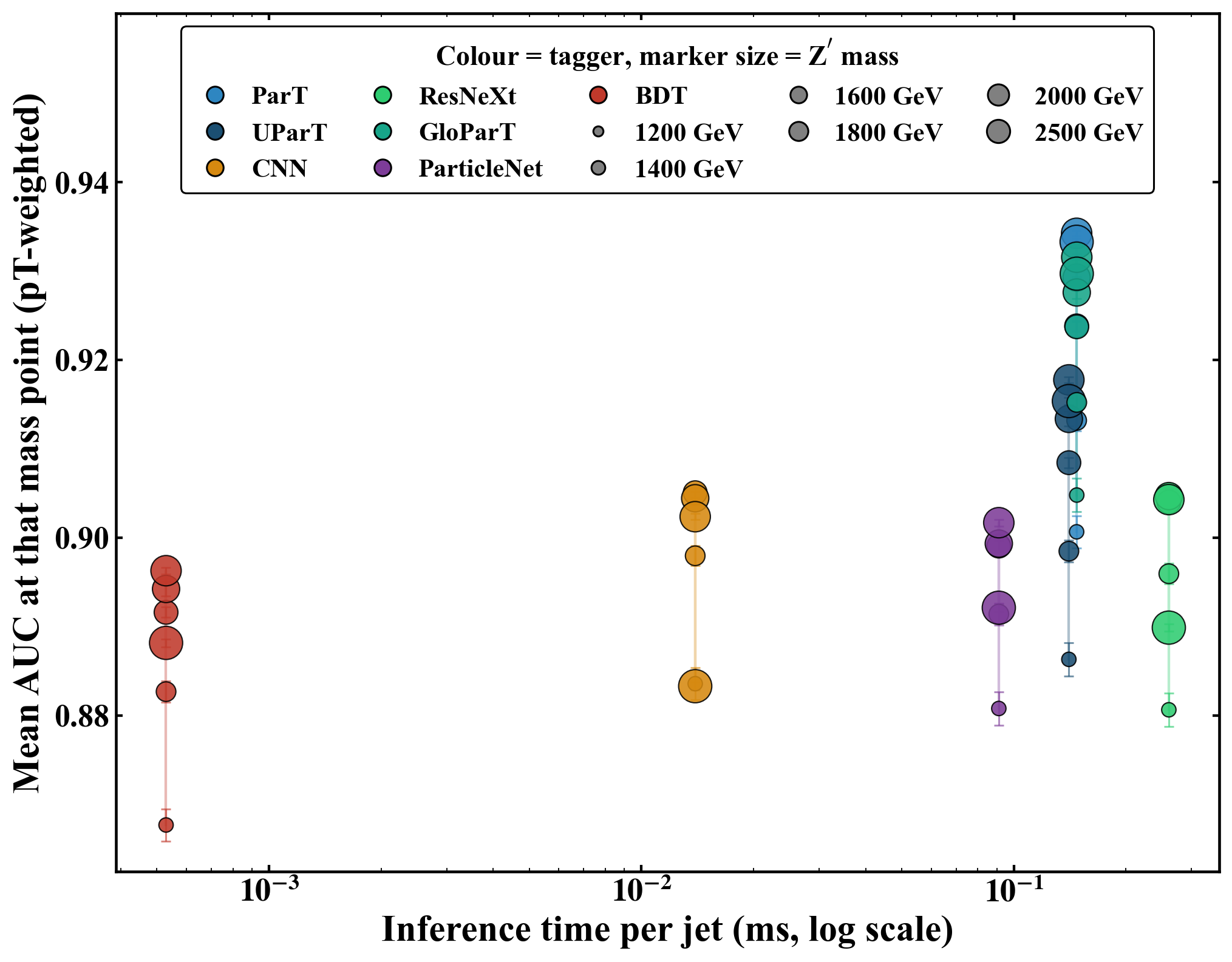}
\caption{}
\label{fig:acc-vs-latency}
\end{subfigure}
\hfill
\begin{subfigure}[b]{0.48\textwidth}
\centering
\includegraphics[width=\textwidth]{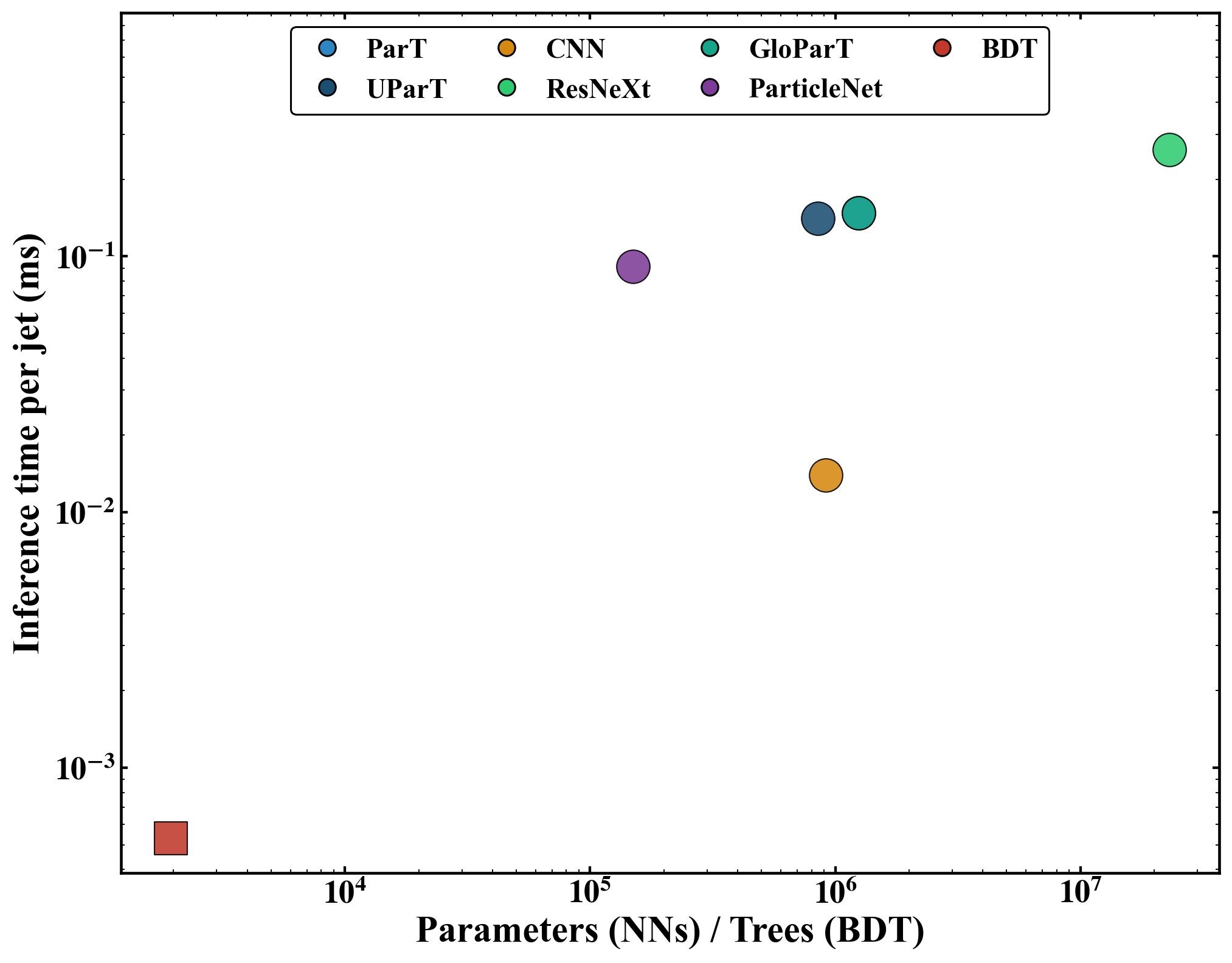}
\caption{}
\label{fig:size-vs-speed}
\end{subfigure}
\caption{Cost and benefit of the seven taggers.
\textbf{(a)} AUC at each $Z'$ mass point vs.\ per-jet inference
latency. \textbf{(b)} Latency vs.\ model size.}
\label{fig:compute-tradeoffs}
\end{figure}

The discrimination of a tagger has to be weighed against its computational cost, especially for the large jet samples of collider analyses. Table~\ref{tab:compute} gives the model size and per-jet inference latency of each tagger measured on an NVIDIA GTX 1080 Ti, and Figure~\ref{fig:compute-tradeoffs} compares these quantities with the AUC. The model size refers to the saved checkpoint on disk or the saved XGBoost model for the BDT. The latency covers only the forward pass of the model and excludes input preparation. This matters for the BDT, which needs subject reclustering, and for the image-based taggers, which need image construction. The computational cost varies widely across the models.

\begin{table}[htbp]
\centering
\caption{Model size is the size of the saved checkpoint on disk.}
\label{tab:compute}
\begin{tabular}{lccc}
\toprule
Model & Parameters / Trees & Model size [MB] & Latency [ms/jet] \\
\midrule
BDT / XGBoost      & 1{,}955 trees   & 24.37 & 0.0010 \\
CNN                & 912{,}833       & 3.49 & 0.0139 \\
ParticleNet        & 149{,}775       & 0.58 & 0.0902 \\
UParT              & 848{,}801       & 3.27 & 0.1410 \\
GloParT            & 1{,}245{,}985   & 4.80 & 0.1468 \\
ParT               & 1{,}246{,}241   & 4.80 & 0.1487 \\
ResNeXt            & 22{,}984{,}002  & 88.04 & 0.2927 \\
\bottomrule
\end{tabular}
\end{table}

The BDT is the fastest tagger, with a latency about 140 times lower than that of transformers, but it also yields the weakest tagging. The CNN runs at about one tenth of the transformer latency, although its mean AUC is lower by $0.015$--$0.032$. Section~\ref{sec:results:significance} showed that such a gap in discrimination can widen at tighter working points. The three transformers have nearly identical latency and model size, as expected from the similar network configurations of ParT and GloParT. More parameters do not guarantee better tagging. ResNeXt has about 25 times as many parameters as the CNN and about 21 times its latency, yet reaches a similar mean AUC. A deeper, larger image network thus yields no gain in discrimination for the chosen image representation. ParticleNet has the smallest model but a latency about 6.5 times that of the CNN. This reflects the extra cost of building neighborhoods and graph edges and agrees with the relative cost reported in the ATLAS comparison~\cite{ATLASAccuracyPrecision}. Tagging performance and computational cost therefore depend on the architecture and its input representation and not on model size alone. The transformers are slower than the BDT and CNN, but they give much stronger discrimination at about $0.15$~ms per jet. This makes them practical candidates for offline analysis.

\subsection{Uncertainties}
\label{sec:results:uncertainty}

\begin{figure}[htbp]
\centering
\includegraphics[width=\textwidth]{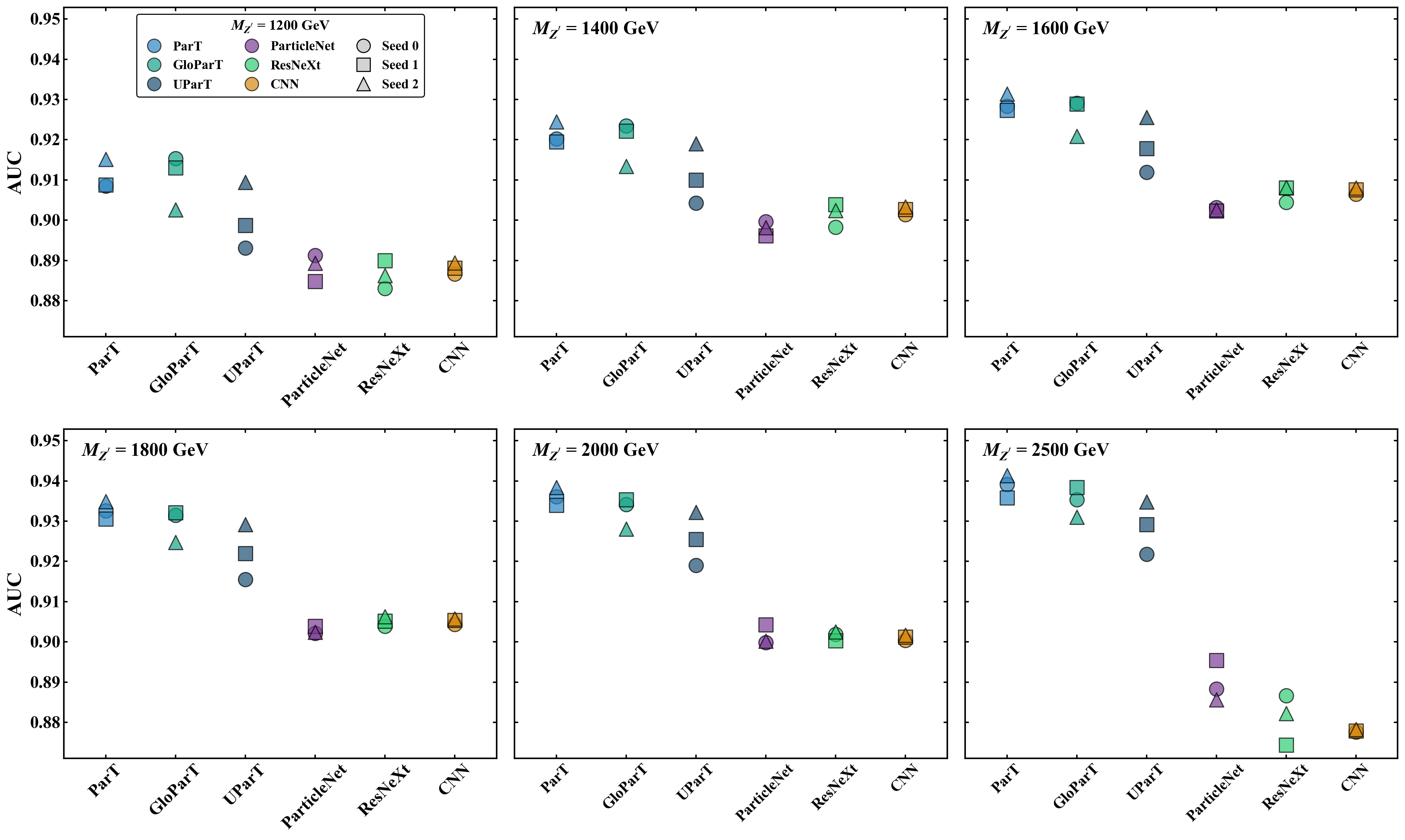}
\caption{AUC of three training seeds of ParT, GloParT, UParT,
ParticleNet, ResNeXt, and the CNN, one panel per $Z'$ mass point.
Colour marks the tagger and marker shape the training seed (seed 0 is
the nominal model of Table~\ref{tab:auc}). The BDT uses a single run.}
\label{fig:seed-variance}
\end{figure}

\textbf{Training-seed variation.} The stability of the tagger ranking under retraining is tested with models trained from different random seeds. Two sources of variation are considered. These are the randomness of the training procedure and the finite size of the simulated samples. The results in Table~\ref{tab:auc} come from a single training run per tagger, so two extra seeds are trained for ParT, UParT, GloParT, ParticleNet, CNN, and ResNeXt. The BDT is not retrained. The same approach is used to assess training-related uncertainties in~\cite{ATLASAccuracyPrecision}. Figure~\ref{fig:seed-variance} shows the resulting AUC values. The gap between the transformers and the other taggers is stable across the three seeds. Within the groups ParT and GloParT sometimes swap places, as do GloParT and UParT, and the differences between ParticleNet and the image-based taggers stay very small. The separation between performance groups is therefore robust, while the exact order inside each group should be read with caution. For this reason, the conclusions of this paper rest on the groups and not on their internal order. For ResNeXt, the results agree across seeds only when each checkpoint is evaluated on images at the resolution used in its training (Section~\ref{sec:reco}).

\textbf{Statistical uncertainties.} All results in Sections~\ref{sec:results:zprime_disc}--\ref{sec:results:compute} come from a finite Monte Carlo sample. To test whether the differences between taggers are statistically significant, a purely statistical $1\sigma$ uncertainty is assigned to each efficiency, AUC, background rejection, and significance and propagated to the relevant figures. The methods account for the per-jet cross-section weights and use analytic expressions, which avoid costly resampling of the large score arrays. The three methods are listed below.
\begin{itemize}
\item Efficiencies are ratios of weighted sums. Their uncertainties are estimated using Clopper--Pearson intervals \cite{ClopperPearson} evaluated on the Kish-effective counts \cite{Kish}, $n_{\mathrm{eff}} = (\sum_i w_i)^2/\sum_i w_i^2$ and $k_{\mathrm{eff}} = \varepsilon n_{\mathrm{eff}}$, with the resulting interval symmetrized to obtain a single uncertainty. This effective-count approximation accommodates small effective counts and efficiencies near zero. The uncertainties on the yields $S$ and $B$ are calculated as $\sqrt{\sum_{i\in\mathrm{passing}} w_i^2}$. If no background jet survives a cut, the one-sided $68.27\%$ upper limit on the background efficiency is quoted, giving a lower limit on the rejection rather than $0\pm0$.
\item The AUC uncertainty is estimated using the weighted DeLong variance \cite{DeLong}. This method also provides the covariance between taggers evaluated on the same jets, allowing the uncertainty on their AUC differences to be calculated.
\item The uncertainties on $Z$ and the background rejection $1/\varepsilon_{\rm bkg}$ are propagated using the derivatives of the corresponding expressions. For the Asimov significance,  the relevant derivatives are 
$\partial Z/\partial S = \ln(1+S/B)/Z$ and 
$\partial Z/\partial B = [\ln(1+S/B)-S/B]/Z$.
\end{itemize}
The uncertainties appear in the figures as annotations in the heatmaps and as bands or error bars on the curves and markers. The differences behind the main conclusions are much smaller than the differences, so the main performance trends are statistically significant. Systematic uncertainties such as those on the jet energy scale and resolution and on the mis-tag rate calibration are not considered here and are left for future work.

\section{Conclusions}
\label{sec:conclusion}

This study tests how the performance of benchmark-trained top taggers carries over to a realistic $Z'\to t\bar{t}$ search without retraining. The overall AUC ranking is largely preserved from JetClass to full CMS simulation (Sections~\ref{sec:results:jetclass} and~\ref{sec:results:zprime_disc}), but the benchmark metrics do not fully reflect the differences in search performance. Taggers with similar AUC differ by up to a factor of $3.3$ in jet-level significance at $99.9\%$ background rejection (Section~\ref{sec:results:significance}). The AUC alone is therefore not enough to choose a tagger for a specific search. Its performance must also be assessed in the high-rejection regime relevant to that analysis. At independently optimized working points, the best transformers reach the jet-level $5\sigma$ threshold up to $1800~\mathrm{GeV}$, compared with $1400~\mathrm{GeV}$ for the BDT (Section~\ref{sec:results:significance}).

A key result is the growing advantage of the particle transformers at higher resonance masses. The comparison restricted to fully merged top jets shows that this advantage does not come only from the change in signal composition with mass (Section~\ref{sec:results:zprime_disc}). The transformers also keep their discrimination better as jet $p_T$ increases, including above the $500\text{--}1000~\mathrm{GeV}$ training range (Section~\ref{sec:results:domainshift}). Holding performance outside the training kinematic range is therefore an important factor when top taggers are applied to searches over a broad mass range. The comparison also sets out the practical trade-off between discrimination and computational cost. ParT and GloParT give the strongest discrimination, while the CNN reaches competitive performance at a much lower inference cost (Section~\ref{sec:results:compute}). The seed and statistical studies support the robustness of the separation between the main performance groups, although the exact order within each group is less certain (Section~\ref{sec:results:uncertainty}). These results provide a quantitative basis for selecting a top tagger based on the kinematic regime, the required background rejection, and the computing limits of the intended analysis.

Top taggers should therefore be evaluated within the physics analysis they are meant to serve. The present study compares their discrimination, stability, and pileup robustness (Section~\ref{subsubsec:pileup}), but several aspects remain open. Systematic uncertainties such as those on the jet energy scale and resolution and on the mis-tag rate calibration are needed for a complete assessment. Further work could extend the training $p_T$ range, match the constituent multiplicities, apply PUPPI pileup mitigation \cite{PUPPI}, add more resonance mass points and fit the full tagger-score distribution. These steps would build on the present results and give a fuller picture of what modern top taggers can achieve in searches for heavy resonances.

\acknowledgments
The corresponding author acknowledges support from the School
of Physical Sciences, IIT Mandi, and from a DST INSPIRE Fellowship.
The authors acknowledge the High-Performance Computing (HPC) facility
at IIT Mandi for the computational resources used in this work.

\bibliographystyle{JHEP}
\bibliography{biblio}

\end{document}

%% file: paper_figures/zprime_illustration_body.tex
\definecolor{topc}{HTML}{C0392B}
\definecolor{wc}{HTML}{2471A3}
\definecolor{bc}{HTML}{1E8449}
\definecolor{qcdc}{HTML}{7D3C98}
\tikzset{lab/.style={font=\large}, ttl/.style={font=\bfseries\large}}

\begin{tikzpicture}[line width=1.1pt, font=\large]

\begin{scope}
\node[ttl,anchor=west] at (-3.4,2.6) {(a)};
\begin{feynman}
  \vertex (q1) at (-3.0, 1.3) {\(q\)};
  \vertex (q2) at (-3.0,-1.3) {\(\bar q\)};
  \vertex (v1) at (-1.6, 0);
  \vertex (v2) at ( 0.2, 0);
  \vertex (t)  at ( 1.5, 1.2);
  \vertex (tb) at ( 1.5,-1.2);
  \vertex (b)  at ( 3.6, 2.3) {\(b\)};
  \vertex (w)  at ( 2.7, 0.9);
  \vertex (wq1) at (3.9, 1.4) {\(q\)};
  \vertex (wq2) at (3.9, 0.35) {\(\bar q'\)};
  \vertex (bb) at ( 3.6,-2.3) {\(\bar b\)};
  \vertex (wm) at ( 2.7,-0.9);
  \vertex (wq3) at (3.9,-0.35) {\(\bar q\)};
  \vertex (wq4) at (3.9,-1.4) {\(q'\)};
  \diagram*{
    (q1) -- [fermion] (v1) -- [fermion] (q2),
    (v1) -- [boson, edge label=\(Z'\)] (v2),
    (v2) -- [fermion, line width=2.2pt, draw=topc, edge label=\(t\)] (t),
    (tb) -- [fermion, line width=2.2pt, draw=topc, edge label'=\(\bar t\)] (v2),
    (t) -- [fermion, draw=bc] (b),
    (t) -- [boson, draw=wc, edge label'=\(W^+\)] (w),
    (w) -- [fermion] (wq1),
    (wq2) -- [fermion] (w),
    (bb) -- [fermion, draw=bc] (tb),
    (tb) -- [boson, draw=wc, edge label=\(W^-\)] (wm),
    (wq3) -- [fermion] (wm),
    (wm) -- [fermion] (wq4),
  };
\end{feynman}
\draw[topc, dashed, line width=1.2pt, rotate around={20:(3.2,1.4)}] (3.2,1.4) ellipse (1.05 and 1.25);
\draw[topc, dashed, line width=1.2pt, rotate around={-20:(3.2,-1.4)}] (3.2,-1.4) ellipse (1.05 and 1.25);
\node[lab, topc, align=center] at (5.2,1.4) {large-$R$\\ top jet};
\node[lab, topc, align=center] at (5.2,-1.4) {large-$R$\\ top jet};
\end{scope}

\begin{scope}[xshift=7.6cm]
\node[ttl,anchor=west] at (-0.8,2.6) {(b)};
\coordinate (O1) at (0,1.5);
\fill (O1) circle (2.5pt);
\draw[-{Latex}, line width=2.2pt, topc] (O1) -- ++(0.9,0) node[above, lab] {$t$};
\coordinate (D1) at (0.9,1.5);
\foreach \ang/\lbl/\col in {30/b/bc, 0/q/wc, -30/{\bar q'}/wc}{
  \draw[-{Latex}, line width=1.4pt, \col] (D1) -- ++(\ang:1.8) coordinate (e);
  \draw[\col, line width=1.2pt, fill=\col!15] (e) circle (0.28);
  \node[lab, \col] at ($(D1)+(\ang:2.35)$) {$\lbl$};
}
\node[lab, align=left, anchor=west] at (3.9,1.5) {resolved\\ $p_T \lesssim 400$~GeV\\ three $R=0.4$ jets};
\coordinate (O2) at (0,-1.5);
\fill (O2) circle (2.5pt);
\draw[-{Latex}, line width=2.2pt, topc] (O2) -- ++(0.9,0) node[above, lab] {$t$};
\coordinate (D2) at (0.9,-1.5);
\draw[topc, dashed, line width=1.2pt, fill=topc!8] ($(D2)+(2.0,0)$) circle (0.75);
\foreach \ang/\lbl/\col in {12/b/bc, 0/q/wc, -12/{\bar q'}/wc}{
  \draw[-{Latex}, line width=1.4pt, \col] (D2) -- ++(\ang:2.25);
}
\node[lab, topc] at ($(D2)+(2.0,-1.0)$) {$R=0.8$};
\node[lab] at ($(D2)+(2.0,1.0)$) {$b\,q\,\bar q'$};
\draw[decorate, decoration={brace, amplitude=3pt}] ($(D2)+(12:2.4)$) -- ($(D2)+(-12:2.4)$)
   node[midway, right=3pt, lab] {$\Delta R \approx 2m_t/p_T$};
\node[lab, align=left, anchor=west] at (3.9,-2.6) {boosted\\ $p_T \gtrsim 430$~GeV\\ one $R=0.8$ jet};
\draw[-{Latex}, line width=1.4pt, gray] (5.6,0.7) -- (5.6,-0.8) node[midway, right, lab, gray, align=left] {higher\\ $m_{Z'}$};
\end{scope}

\end{tikzpicture}